\documentclass[fleqn,usenatbib,useAMS]{mnras}
 \usepackage{graphicx}
 \usepackage{amsmath}
 \usepackage{amssymb}
 \usepackage[T1]{fontenc}
 \usepackage{ae,aecompl}
 \usepackage{txfonts}
 \usepackage{enumerate}

 \title[Orbital evolution of 2017~FZ$_{2}$ et al.]
       {Asteroid 2017~FZ$_\mathbf{2}$ et al.: signs of recent mass-shedding from YORP?}

 \author[C. de la Fuente Marcos and R. de la Fuente Marcos]
        {C.~de~la~Fuente~Marcos\thanks{E-mail: nbplanet@ucm.es}
         and
         R. de la Fuente Marcos \\
         Universidad Complutense de Madrid,
         Ciudad Universitaria, E-28040 Madrid, Spain}
 \date{Accepted 2017 September 25. 
       Received 2017 September 25; 
       in original form 2017 June 7}
 \pubyear{2017}
 
\begin{document}
  \label{firstpage}
  \pagerange{\pageref{firstpage}--\pageref{lastpage}}
  \maketitle

  \begin{abstract}
     The first direct detection of the asteroidal YORP effect, a phenomenon 
     that changes the spin states of small bodies due to thermal reemission 
     of sunlight from their surfaces, was obtained for (54509)~YORP 
     2000~PH$_{5}$. Such an alteration can slowly increase the rotation rate 
     of asteroids, driving them to reach their fission limit and causing 
     their disruption. This process can produce binaries and unbound asteroid 
     pairs. Secondary fission opens the door to the eventual formation of 
     transient but genetically-related groupings. Here, we show that the 
     small near-Earth asteroid (NEA) 2017~FZ$_{2}$ was a co-orbital of our 
     planet of the quasi-satellite type prior to their close encounter on 
     2017 March 23. Because of this flyby with the Earth, 2017~FZ$_{2}$ has 
     become a non-resonant NEA. Our $N$-body simulations indicate that this 
     object may have experienced quasi-satellite engagements with our planet 
     in the past and it may return as a co-orbital in the future. We identify 
     a number of NEAs that follow similar paths, the largest named being 
     YORP, which is also an Earth's co-orbital. An apparent excess of NEAs 
     moving in these peculiar orbits is studied within the framework of two 
     orbit population models. A possibility that emerges from this analysis 
     is that such an excess, if real, could be the result of mass shedding 
     from YORP itself or a putative larger object that produced YORP. Future 
     spectroscopic observations of 2017~FZ$_{2}$ during its next visit in 
     2018 (and of related objects when feasible) may be able to confirm or 
     reject this interpretation.
  \end{abstract}


  \section{Introduction}
     Small near-Earth asteroids (NEAs) are interesting targets because their study can lead to a better understanding of the evolution of 
     the populations of larger minor bodies from which they originate. Most large asteroids consist of many types of rocks held together by
     gravity and friction between fragments; in stark contrast, most small asteroids are thought to be fast-spinning bare boulders (see e.g.
     Harris 2013; Statler et al. 2013; Hatch \& Wiegert 2015; Polishook et al. 2015; Ryan \& Ryan 2016). Small NEAs can be chipped off by 
     another small body from a larger parent asteroid through subcatastrophic impacts (see e.g. Durda et al. 2007), they can also be 
     released during very close encounters with planets following tidal disruption (see e.g. Keane \& Matsuyama 2014; Schunov{\'a} et al. 
     2014), or due to the action of the Yarkovsky--O'Keefe--Radzievskii--Paddack (YORP) mechanism (see e.g. Bottke et al. 2006). 

     The asteroidal YORP effect changes the spin states of small bodies as a result of thermal reemission of starlight from their surfaces. 
     Such an alteration can secularly increase the rotation rate of asteroids, driving them to reach their fission limit and subsequently 
     triggering their disruption (Walsh, Richardson \& Michel 2008). This process can produce binary systems and unbound asteroid pairs 
     (Vokrouhlick{\'y} \& Nesvorn{\'y} 2008; Pravec et al. 2010; Scheeres 2017). Asteroids formed by rotational fission escape from each 
     other if the size of one of the members of the pair is small enough (see e.g. Jacobson \& Scheeres 2011). Secondary fission opens the 
     door to the formation of transient, but genetically-related, dynamic groupings. 

     The discovery and ensuing study of the orbital evolution of (54509)~YORP 2000~PH$_{5}$ led to the first direct observational detection 
     of the YORP effect (Lowry et al. 2007; Taylor et al. 2007). While YORP spin-up is now widely considered as the dominant formation 
     mechanism for small NEAs (see e.g. Walsh et al. 2008; Jacobson \& Scheeres 2011; Walsh, Richardson \& Michel 2012; Jacobson et al. 
     2016), there are still major questions remaining to be answered. In particular, meteoroid impacts can also affect asteroid spins at a 
     level comparable to that of the YORP effect under certain circumstances (Henych \& Pravec 2013; Wiegert 2015). In fact, it has been 
     suggested that small asteroids can be used to deflect incoming NEAs via kinetic impacts (Akiyama, Bando \& Hokamoto 2016).

     The asteroidal YORP effect has been measured in objects other than YORP; for example, (25143)~Itokawa 1998~SF$_{36}$ (Kitazato et al. 
     2007; {\v D}urech et al. 2008a; Lowry et al. 2014), (1620)~Geographos 1951~RA ({\v D}urech et al. 2008b), (3103)~Eger 1982~BB 
     ({\v D}urech et al. 2012) and small asteroids in the Karin cluster (Carruba, Nesvorn{\'y} \& Vokrouhlick{\'y} 2016). P/2013 R3, a 
     recent case of asteroid breakup, has been found to be consistent with the YORP-induced rotational disruption of a weakly bound minor 
     body (Jewitt et al. 2017). On the other hand, YORP is not only known for being affected by the YORP effect, it is also a well-studied 
     co-orbital of the Earth (Wiegert et al. 2002; Margot \& Nicholson 2003). Members of this peculiar dynamical class are subjected to the 
     1:1 mean-motion resonance with our planet. 

     Earth's co-orbital zone currently goes from $\sim$0.994~au to $\sim$1.006~au, equivalent to a range in orbital periods of 362 to 368~d 
     (see e.g. de la Fuente Marcos \& de la Fuente Marcos 2016f). Models (see Section 3.2) indicate that the probability of a NEA ever 
     ending up on an orbit within this region has an average value of nearly 0.0025. Such an estimate matches the observational results well 
     (see Section 3.2), although NEAs in Earth's co-orbital zone can only be observed for relatively brief periods of time due to their long 
     synodic periods. About 65 per cent of all known NEAs temporarily confined or just passing through Earth's co-orbital zone have absolute 
     magnitude, $H$, greater than 22~mag, or a size of the order of 140~m or smaller, making them obvious candidates to being by-products of 
     YORP spin-up or perhaps other processes capable of triggering fragmentation events (see above). 

     Here, we show that the recently discovered minor body 2017~FZ$_{2}$ was until very recently a quasi-satellite of the Earth and argue 
     that it could be related to YORP, which is also a transient companion to the Earth moving in a horseshoe-type orbit (Wiegert et al. 
     2002; Margot \& Nicholson 2003). This paper is organized as follows. In Section 2, we present data, details of our numerical model, and 
     2017~FZ$_{2}$'s orbital evolution. Section 3 explores the possibility of the existence of a dynamical grouping, perhaps related to 
     YORP. Mutual close encounters between members of this group are studied in Section 4. Close approaches to other NEAs, Venus, the Earth 
     and Mars are investigated in Section 5. Our results are discussed in Section 6. Section 7 summarizes our conclusions.    

  \section{Asteroid 2017~FZ$_{2}$: data, integrations and orbital evolution}
     The recently discovered NEA 2017~FZ$_{2}$ was originally identified as an Earth's co-orbital candidate because of its small relative 
     semimajor axis, $|a-a_{\rm Earth}|\sim$0.0012~au, at discovery time (now it is 0.008~au). Here, we present the data currently available 
     for this object, outline the techniques used in its study, and explore both its short- and medium-term orbital evolution.

     \subsection{The data}
        Asteroid 2017~FZ$_{2}$ was discovered on 2017 March 19 by G.~J.~Leonard observing with the 1.5-m reflector telescope of the Mt. 
        Lemmon Survey at an apparent magnitude $V$ of 19.2 (Urakawa et al. 2017).\footnote{http://www.minorplanetcenter.net/mpec/K17/K17F65.html} 
        It subsequently made a close approach to our planet on 2017 March 23 when it came within a nominal distance of 0.0044~au, travelling  
        at a relative velocity of 8.46~km~s$^{-1}$.\footnote{https://ssd.jpl.nasa.gov/sbdb.cgi?sstr=2017\%20FZ2;old=0;orb=0;\\cov=0;log=0;cad=1\#cad}
        The object was observed with radar from Arecibo Observatory (Rivera-Valentin et al. 2016) on 2017 March 27, when it was receding 
        from our planet.\footnote{http://www.naic.edu/\%7Epradar/asteroids/2017FZ2/2017FZ2.2017Mar\\27.s2p0Hz.cw.png} The orbital solution 
        currently available for this object (see Table \ref{elements}) is based on 152 observations spanning a data-arc of 8~d (including 
        the Doppler observation); its minimum orbit intersection distance (MOID) with the Earth is 0.0014~au. 

        Asteroid 2017~FZ$_{2}$ was initially included in the list of asteroids that may be involved in potential future Earth impact events 
        compiled by the Jet Propulsion Laboratory (JPL) Sentry System (Chamberlin et al. 2001; Chodas 2015),\footnote{https://cneos.jpl.nasa.gov/sentry/} 
        with a computed impact probability of 0.000092 for a possible impact in 2101--2104, but it has been removed 
        since.\footnote{https://cneos.jpl.nasa.gov/sentry/details.html\#?des=2017\%20FZ2} It is, however, a small object with $H=26.7$~mag 
        (assumed $G$ = 0.15), which suggests a diameter in the range 13--30~m for an assumed albedo in the range 0.20--0.04. For this 
        reason, the explosive energy associated with a hypothetical future impact of this minor body could be comparable to those of typical 
        nuclear weapons currently stocked, i.e. a locally dangerous impact not too different from that of the Chelyabinsk event (see e.g. 
        Brown et al. 2013; Popova et al. 2013; de la Fuente Marcos \& de la Fuente Marcos 2015c).  
%
%
     \begin{table*}
        \fontsize{8}{12pt}\selectfont
        \tabcolsep 0.30truecm
        \caption{Heliocentric Keplerian orbital elements and 1$\sigma$ uncertainties of 2017~FZ$_{2}$, 2017~DR$_{109}$ and (54509)~YORP 
                 2000~PH$_{5}$. The orbital solutions have been computed at epoch JD 2458000.5 that corresponds to 00:00:00.000 TDB, 
                 Barycentric Dynamical Time, on 2017 September 4 (J2000.0 ecliptic and equinox. Source: JPL's Small-Body Database.)
                }
        \centering
        \begin{tabular}{lcccc}
           \hline
            Orbital parameter                                 &   & 2017~FZ$_{2}$           & 2017~DR$_{109}$     & YORP                          \\ 
           \hline
            Semimajor axis, $a$ (au)                          & = & 1.0071385$\pm$0.0000007 & 1.00064$\pm$0.00007 & 1.0062210205$\pm$0.0000000003 \\
            Eccentricity, $e$                                 & = & 0.264054$\pm$0.000002   & 0.2414$\pm$0.0003   & 0.23021775$\pm$0.00000014     \\
            Inclination, $i$ (\degr)                          & = & 1.81167$\pm$0.00002     & 3.060$\pm$0.004     & 1.599312$\pm$0.000005         \\
            Longitude of the ascending node, $\Omega$ (\degr) & = & 185.86918$\pm$0.00002   & 341.3111$\pm$0.0005 & 278.28130$\pm$0.00007         \\
            Argument of perihelion, $\omega$ (\degr)          & = & 100.32304$\pm$0.00009   & 72.094$\pm$0.003    & 278.86596$\pm$0.00006         \\
            Mean anomaly, $M$ (\degr)                         & = &  87.30597$\pm$0.00009   & 263.04$\pm$0.05     &  79.660894$\pm$0.000012       \\
            Perihelion, $q$ (au)                              & = & 0.741200$\pm$0.000002   & 0.7591$\pm$0.0002   & 0.77457108$\pm$0.00000014     \\
            Aphelion, $Q$ (au)                                & = & 1.2730773$\pm$0.0000008 & 1.24218$\pm$0.00009 & 1.2378709621$\pm$0.0000000003 \\
            Absolute magnitude, $H$ (mag)                     & = & 26.7$\pm$0.4            & 27.6$\pm$0.5        & 22.7                          \\
           \hline
        \end{tabular}
        \label{elements}
     \end{table*}
%
%

     \subsection{The approach}
        As explained in de la Fuente Marcos \& de la Fuente Marcos (2016f), confirmation of co-orbital candidates of a given host comes only 
        after the statistical analysis of the behaviour of a critical or resonant angle, $\lambda_{\rm r}$,\footnote{In our case, the 
        relative mean longitude or difference between the mean longitude of the object and that of its host.} in a relevant set of numerical 
        simulations that accounts for the uncertainties associated with the orbit determination of the candidate. If the value of 
        $\lambda_{\rm r}$ librates or oscillates over time, the object is actually trapped in a 1:1 mean-motion resonance with its host as 
        their orbital periods are virtually the same; if $\lambda_{\rm r}$ circulates in the interval (0, 360)\degr, we speak of a 
        non-resonant, passing body. Librations about 0{\degr} (quasi-satellite), $\pm$60{\degr} (Trojan) or 180{\degr} (horseshoe) are often 
        cited as the signposts of 1:1 resonant behaviour (see e.g. Murray \& Dermott 1999), although hybrids of the three elementary 
        co-orbital states are possible and the actual average resonant value of $\lambda_{\rm r}$ depends on the orbital eccentricity and 
        inclination of the object (Namouni, Christou \& Murray 1999; Namouni \& Murray 2000). 

        Here, we use a direct $N$-body code\footnote{http://www.ast.cam.ac.uk/\%7Esverre/web/pages/nbody.htm} implemented by Aarseth (2003) 
        and based on the Hermite scheme described by Makino (1991) ---i.e. no linear or non-linear secular theory is used in this study--- 
        to investigate the orbital evolution of 2017~FZ$_{2}$ and several other, perhaps related, NEAs. The results of Solar system 
        calculations performed with this code are consistent with those obtained by other authors using different softwares (see de la 
        Fuente Marcos \& de la Fuente Marcos 2012); for further details, including the assumed physical model, see de la Fuente Marcos \& de 
        la Fuente Marcos (2012, 2016f). Initial conditions in the form of positions and velocities in the barycentre of the Solar system for 
        2017~FZ$_{2}$, other relevant NEAs, and the various bodies that define the physical model have been obtained from JPL's 
        \textsc{horizons}\footnote{https://ssd.jpl.nasa.gov/?horizons} system (Giorgini et al. 1996; Standish 1998; Giorgini \& Yeomans 
        1999; Giorgini, Chodas \& Yeomans 2001; Giorgini 2011, 2015) at epoch JD 2458000.5 (2017-September-04.0 TDB, Barycentric Dynamical 
        Time), which is the $t$ = 0 instant in our figures unless explicitly stated.

     \subsection{The evolution}
\begin{figure}
        \centering
        \includegraphics[width=\linewidth]{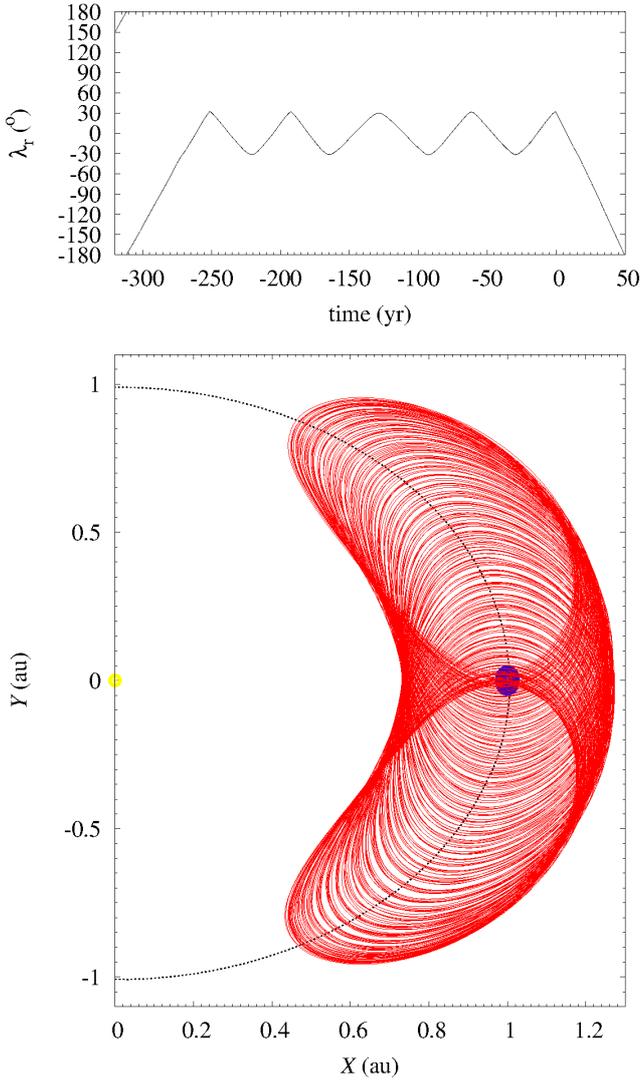}
        \caption{Most recent quasi-satellite episode of 2017~FZ$_{2}$. The top panel shows the behaviour of the resonant angle, 
                 $\lambda_{\rm r}$; the bottom panel displays the path followed by this minor body in a frame of reference centred at the 
                 Sun and rotating with our planet, projected on to the ecliptic plane. The diagram also includes the orbit of the Earth, 
                 its position at (1, 0)~au, and the Sun at (0, 0)~au. The figure shows the evolution of the nominal orbital solution as in 
                 Table \ref{elements}, left-hand column, but the behaviour observed around $t=0$ is common to all investigated control 
                 orbits.  
                }
        \label{orbit}
     \end{figure}

        Fig. \ref{orbit}, top panel, shows that, prior to its close encounter with our planet on 2017 March 23, 2017~FZ$_{2}$ was a 
        quasi-satellite of the Earth with a period close to 60~yr as the value of the resonant angle was librating about zero with an 
        amplitude of nearly 30\degr. The bottom panel in Fig. \ref{orbit} only displays the time interval ($-$250, 1)~yr for clarity and 
        shows complex, drifting yearly loops (the annual epicycles) as seen in a frame of reference centred at the Sun and rotating with the 
        Earth. This result ---for the time interval ($-$225, 50)~yr--- is statistically robust as it is common to all the control orbits 
        (over 10$^{3}$) investigated in this work. Extensive calculations (see below) show that the orbital evolution of this NEA is highly 
        sensitive to initial conditions, much more than for any other previously documented quasi-satellite of our planet. A very chaotic 
        orbit implies that it will be difficult to reconstruct its past dynamical evolution or make reliable predictions about its future 
        behaviour beyond a few hundred years.
%
%
%
%

\begin{figure}
        \centering
        \includegraphics[width=\linewidth]{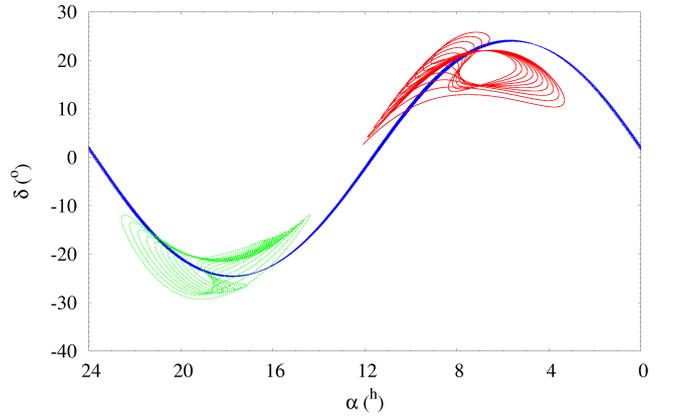}
        \caption{Apparent motion in geocentric equatorial coordinates of 2009~HE$_{60}$ (green), 2017~DR$_{109}$ (blue) and 2017~FZ$_{2}$ 
                 (red) over the time interval ($-$25, $-$15) yr. The three objects were moving co-orbital to the Earth during the displayed 
                 time window, but 2009~HE$_{60}$ and 2017~FZ$_{2}$ were quasi-satellites and 2017~DR$_{109}$ was following a horseshoe-type 
                 path.
                }
        \label{qsloops}
     \end{figure}

        When observed from the ground, a quasi-satellite of the Earth traces an analemma in the sky (de la Fuente Marcos \& de la Fuente 
        Marcos 2016e,f).\footnote{There is an error in terms of quoted units in figs 1 and 2 of de la Fuente Marcos \& de la Fuente Marcos 
        (2016e), and figs 2 and 3 of de la Fuente Marcos \& de la Fuente Marcos (2016f), the right ascension is measured in hours not 
        degrees in those figures.} Fig. \ref{qsloops} shows ten loops of the analemmatic curve described by 2017~FZ$_{2}$ (in red, nominal 
        orbit) that is the result from the interplay between the tilt of the rotational axis of the Earth and the properties of the orbit of 
        the quasi-satellite. Due to its significant eccentricity but low orbital inclination, its apparent motion traces a very distorted 
        teardrop. Non-quasi-satellite co-orbitals do not trace analemmatic loops as seen from the Earth (see the blue curve in 
        Fig.~\ref{qsloops} that corresponds to 2017~DR$_{109}$, an Earth's co-orbital that follows a horseshoe-type path).
%
%
%
%

        Consequently, 2017~FZ$_{2}$ joins the list of quasi-satellites of our planet that already includes (164207) 2004~GU$_{9}$ (Connors 
        et al. 2004; Mikkola et al. 2006; Wajer 2010), (277810) 2006~FV$_{35}$ (Wiegert et al. 2008; Wajer 2010), 2013~LX$_{28}$ (Connors 
        2014), 2014~OL$_{339}$ (de la Fuente Marcos \& de la Fuente Marcos 2014, 2016c) and (469219) 2016~HO$_{3}$ (de la Fuente Marcos \& 
        de la Fuente Marcos 2016f). Although it was the smallest known member of the quasi-satellite dynamical class ---independent of the 
        host planet and significantly smaller than the previous record holder, 469219, that has $H$ = 24.2 mag--- it is no longer a member 
        of this category; our calculations indicate that, after its most recent close flyby with our planet, it has become a non-resonant 
        NEA.

        The quasi-satellite resonant state ---as the one experienced by 2017~FZ$_{2}$ and the other five objects pointed out above--- was 
        first described theoretically by Jackson (1913), but without the use of modern terminology. The energy balance associated with it 
        was studied by H\'enon (1969), who called the objects engaged in this unusual resonant behaviour `retrograde satellites'. However, 
        such objects are not true satellites because they are not gravitationally bound to a host (i.e. have positive planetocentric 
        energy). The term `quasi-satellite' itself was first used in its present sense by Mikkola \& Innanen (1997). Although the first 
        quasi-satellite (of Jupiter) may have been identified (and lost) in 1973 (Chebotarev 1974),\footnote{Originally published in 
        Russian, Astron. Zh., 50, 1071-1075 (1973 September--October).} the first bona fide quasi-satellite (of Venus in this case), 
        2002~VE$_{68}$, was documented by Mikkola et al. (2004). A modern theory of quasi-satellites has been developed in the papers by 
        Mikkola et al. (2006), Sidorenko et al. (2014) and Pousse, Robutel \& Vienne (2017). A recent review of confirmed quasi-satellites 
        has been presented by de la Fuente Marcos \& de la Fuente Marcos (2016e).  

\begin{figure*}
        \centering
        \includegraphics[width=\linewidth]{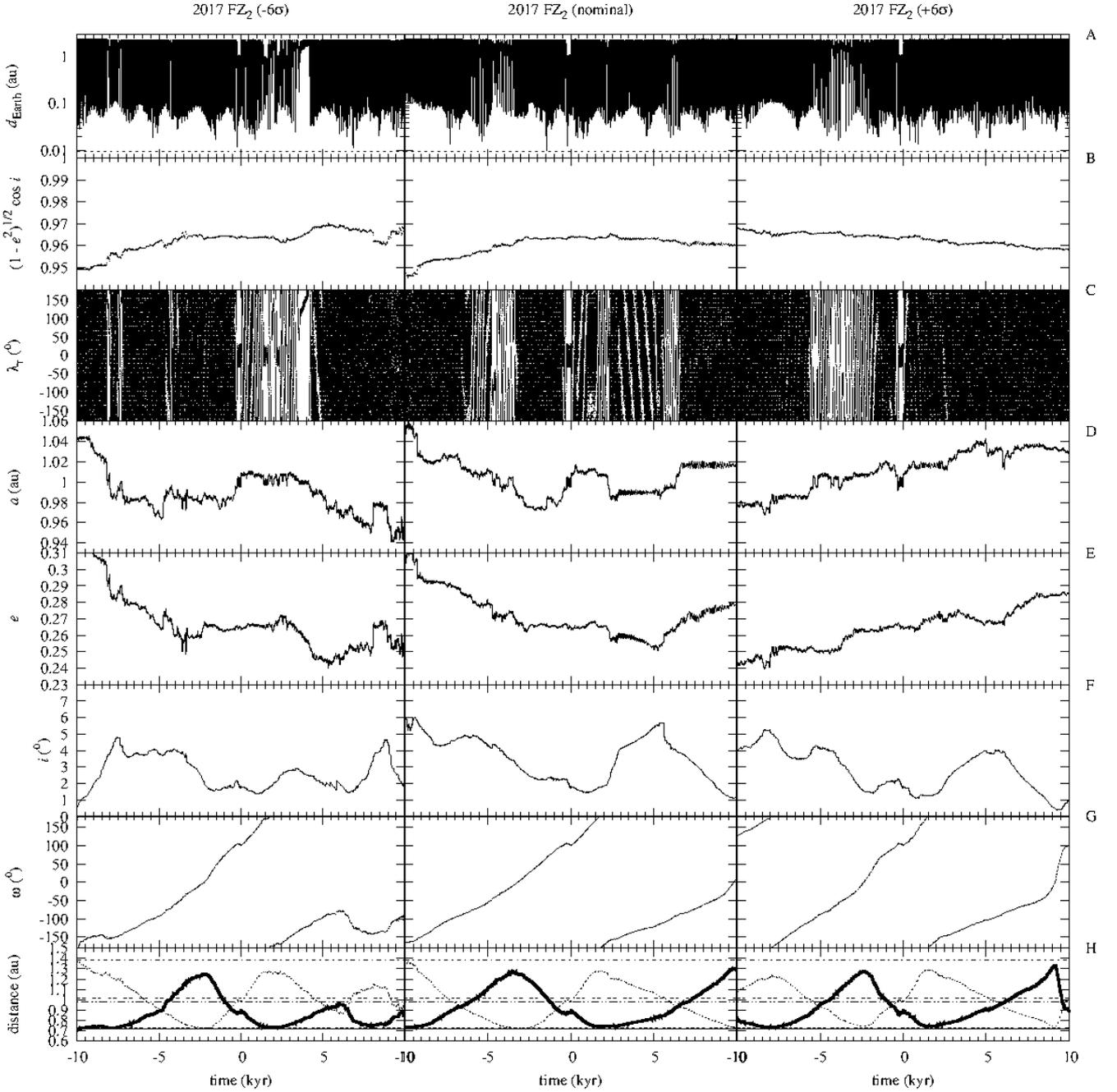}
        \caption{Evolution over time of the values of the orbital elements and other relevant parameters for the nominal orbit of 
                 2017~FZ$_{2}$ as in Table \ref{elements}, left-hand column (central panels), and two representative examples of control 
                 orbits that are very different from the nominal one (but still marginally compatible with the observations, see the text 
                 for details). The A-panels show the evolution of the geocentric distance; the value of the Hill radius of the Earth, 
                 0.0098~au, is also plotted as reference. The B-panels focus on the evolution of the value of the Kozai-Lidov parameter. The 
                 C-panels show the value of the resonant angle. The D-panels, E-panels, F-panels and G-panels show respectively the 
                 evolution of the values of semimajor axis, eccentricity, inclination and argument of perihelion of the control orbits. In 
                 the H-panels, the distances from the descending (thick line) and ascending nodes (dotted line) to the Sun are displayed; 
                 Earth's, Venus' and Mars' aphelion and perihelion distances are shown as well.
                }
        \label{control}
     \end{figure*}

        Fig. \ref{orbit} shows that the most recent quasi-satellite episode of 2017~FZ$_{2}$ started nearly 275~yr ago (but see below) and 
        ended after a close encounter with the Earth on 2017 March 23. The past and future orbital evolution of this object as described by 
        its nominal orbit in Table~\ref{elements}, left-hand column, is shown in Fig.~\ref{control}, central panels; the evolution of two 
        representative control orbits based on the nominal solution but adding (+) or subtracting ($-$) six times the corresponding 
        uncertainty from each orbital element (all the six parameters) in Table~\ref{elements}, left-hand column, and labelled as 
        `$\pm6\sigma$' are displayed as well (right-hand and left-hand panels, respectively). These two examples of orbit evolution using 
        initial conditions that are most different from those associated with the nominal orbit are not meant to show how wide the 
        dispersion of the various parameters could be as they change over time; the statistical effect of the uncertainties will be studied 
        later. 

        Figs \ref{orbit} and \ref{control} show that prior to its most recent flyby with our planet this NEA was an Aten asteroid (now it is 
        an Apollo) following a moderately eccentric orbit, $e$ = 0.26, with low inclination, $i$ = 1\fdg71, that kept the motion of this 
        minor body confined between the orbits of Venus and Mars as it experienced close approaches to both Venus and the Earth (A- and 
        H-panels). These two planets are the main direct perturbers of 2017~FZ$_{2}$ ---although Jupiter drives the precession of the nodes 
        (see e.g. Wiegert, Innanen \& Mikkola 1998). For this reason, the dynamical context of this NEA is rather different from that of the 
        recently identified Earth's quasi-satellite 469219 (de la Fuente Marcos \& de la Fuente Marcos 2016f). The geocentric distance 
        during the closest approaches shown in Fig.~\ref{control}, A-panels, is often lower than the one presented in the figure due to its 
        limited time resolution (a data output interval of 36.5~d was used for these calculations); for example, our (higher time 
        resolution) calculations show that on 2017 March 23 the minimum distance between 2017~FZ$_{2}$ and our planet was about 0.0045~au 
        although Fig.~\ref{control}, central A-panel, shows a value well above the Hill radius of the Earth, 0.0098~au, during the 
        encounter.
%
%
%
%

        Fig. \ref{control} shows that 2017~FZ$_{2}$ experienced brief quasi-satellite engagements with our planet in the past and it will 
        return as Earth's co-orbital in the future (C-panels); it may remain inside or close to the neighbourhood of Earth's co-orbital zone 
        for 20~kyr and possibly more although its orbital evolution is very chaotic (see below). The value of the Kozai-Lidov parameter 
        $\sqrt{1 - e^2} \cos i$ (Kozai 1962; Lidov 1962) remains fairly constant (B-panels). In our case, oscillation of the argument of 
        perihelion is observed for a certain period of time (Fig.~\ref{control}) at $\omega$ = 270{\degr} (left-hand G-panels); libration 
        about 180{\degr} has also been observed for other control orbits, but it is not shown here. When $\omega$ oscillates about 
        180{\degr} the NEA reaches perihelion while approaching the descending node; when $\omega$ librates about 270{\degr} ($-$90\degr), 
        aphelion always occurs away from the orbital plane of the Earth (and perihelion away from Venus). Some of these episodes correspond 
        to domain III evolution as described in Namouni (1999), i.e. horseshoe-retrograde satellite orbit transitions and librations. This 
        behaviour is also observed for 469219 (de la Fuente Marcos \& de la Fuente Marcos 2016f) and other Earth's co-orbitals or near 
        co-orbitals (de la Fuente Marcos \& de la Fuente Marcos 2015b, 2016a,b).
%
%
     \begin{figure}
        \centering
        \includegraphics[width=\linewidth]{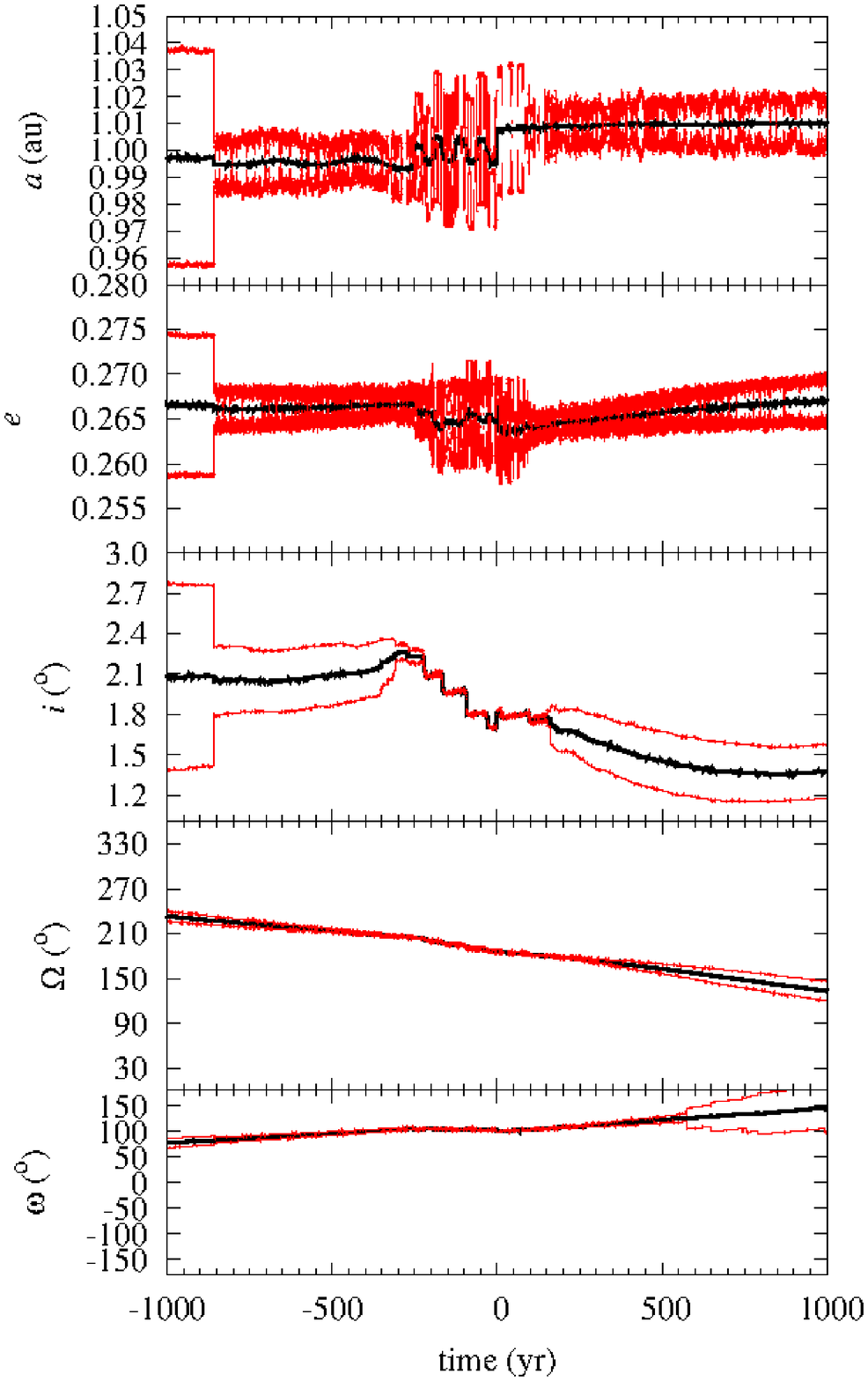}
        \caption{Time evolution of the dispersions of the values of the orbital elements of 2017~FZ$_{2}$: semimajor axis (top panel), 
                 eccentricity (second to top panel), inclination (middle panel), longitude of the ascending node (second to bottom panel), 
                 and argument of perihelion (bottom panel). Average values are displayed as thick black curves and their ranges (1$\sigma$ 
                 uncertainties) as thin red curves. The various panels show results for 250 control orbits; initial positions and velocities 
                 for 2017~FZ$_{2}$ have been computed as explained in section 3 of de la Fuente Marcos \& de la Fuente Marcos (2015d), using 
                 the covariance matrix.
                }
        \label{disper}
     \end{figure}
%
%

        The current orbital solution for 2017~FZ$_{2}$ is not as good as that of 469219, the fifth quasi-satellite of our planet, which has 
        a data-arc spanning 13.19~yr; in any case, the overall evolution of 2017~FZ$_{2}$ is significantly more chaotic as it is subjected 
        to the direct perturbation of Venus and the Earth--Moon system. The combination of relatively poor orbital determination and strong 
        direct planetary perturbations makes it difficult to predict the dynamical status of this object beyond a few centuries. Following 
        the approach detailed in de la Fuente Marcos \& de la Fuente Marcos (2015d), we have applied the Monte Carlo using the Covariance 
        Matrix (MCCM) method to investigate the impact of the uncertainty in the orbit of 2017~FZ$_{2}$ on predictions of its past and 
        future evolution. Here and elsewhere in this paper, covariance matrices have been obtained from JPL's \textsc{horizons}. 

\begin{figure}
        \centering
        \includegraphics[width=\linewidth]{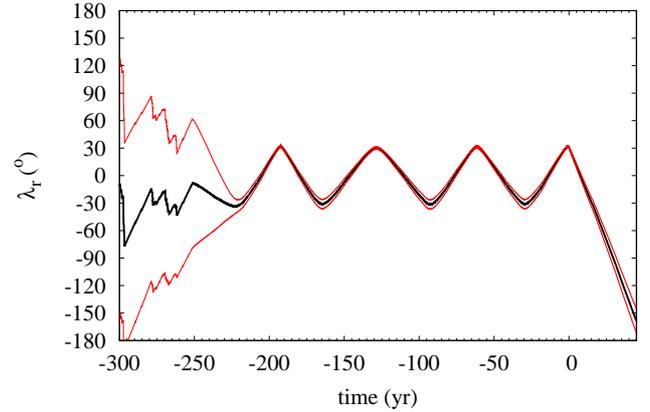}
        \caption{Time evolution of the dispersion of the value of the resonant angle, $\lambda_{\rm r}$, of 2017~FZ$_{2}$. The average value 
                 is displayed as a thick black curve and its ranges (1$\sigma$ uncertainty) as thin red curves. This figure shows results 
                 for 250 control orbits; initial positions and velocities for 2017~FZ$_{2}$ have been computed as explained in section 3 of 
                 de la Fuente Marcos \& de la Fuente Marcos (2015d), using the covariance matrix.
                }
        \label{QSdisper}
     \end{figure}

        Fig.~\ref{disper} shows results for 250 control orbits. Its future evolution is rather uncertain, which is typical of minor bodies 
        with a perhaps non-negligible probability of colliding with our planet during the next century or so ---as pointed out above, an 
        impact was thought to be possible in 2101--2104 and the dispersion grows very significantly after that. In the computation of this 
        set of control orbits (their orbital elements), the Box-Muller method (Box \& Muller 1958; Press et al. 2007) has been used to 
        generate random numbers according to the standard normal distribution with mean 0 and standard deviation 1 (for additional details, 
        see de la Fuente Marcos \& de la Fuente Marcos 2015d, 2016f). Our calculations show that the orbit of 2017~FZ$_{2}$ is inherently 
        very unstable due to its close planetary flybys and has a Lyapunov time ---or characteristic time-scale for exponential divergence 
        of integrated orbits that start arbitrarily close to each other--- of the order of 10$^{2}$~yr. Such short values of the Lyapunov 
        time are typical of planet-crossing co-orbitals (see e.g. Wiegert et al. 1998). Fig.~\ref{QSdisper} clearly shows how this 
        circumstance affects the value of the resonant angle and the quasi-satellite nature of 2017~FZ$_{2}$ (compare with Fig. \ref{orbit}, 
        top panel).
%
%
%
%

        Asteroid 2017~FZ$_{2}$ will reach its next visibility window for observations from the ground starting in February 2018. From 
        February 27 to March 18, it will be observable at an apparent visual magnitude $\leq23$ cruising from right ascension 10$^{\rm h}$ 
        to 8$^{\rm h}$ and declination +1{\degr} to +14\degr. Unfortunately, the Moon will interfere with any planned observations until 
        March 10. This will be the best opportunity to gather spectroscopy of this object for the foreseeable future.    

\begin{figure}
        \centering
        \includegraphics[width=\linewidth]{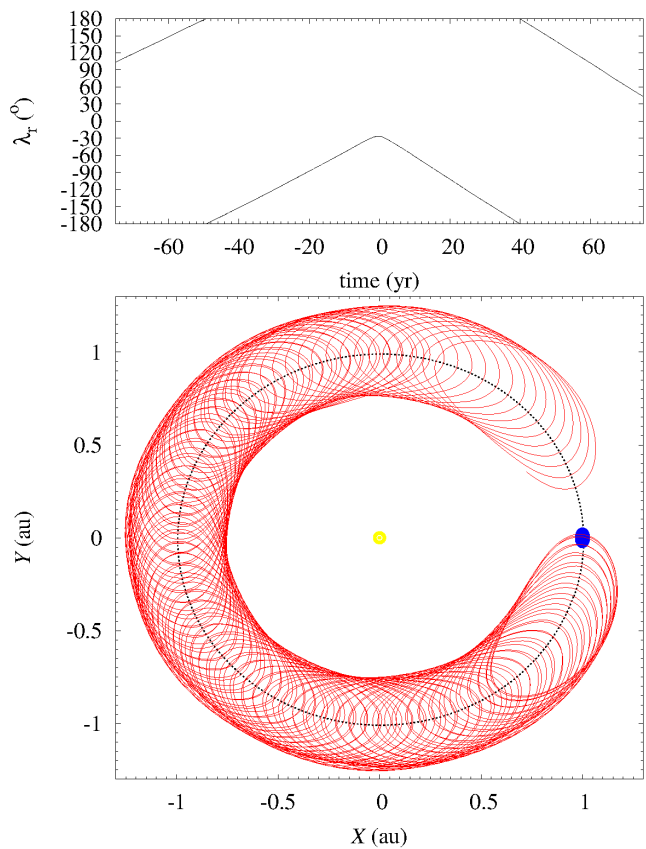}
        \caption{Similar to Fig. \ref{orbit} but for 2017~DR$_{109}$. For clarity, only the time interval ($-$75, 75)~yr is displayed.
                }
        \label{orbitDR109}
     \end{figure}

\begin{figure*}
        \centering
        \includegraphics[width=\linewidth]{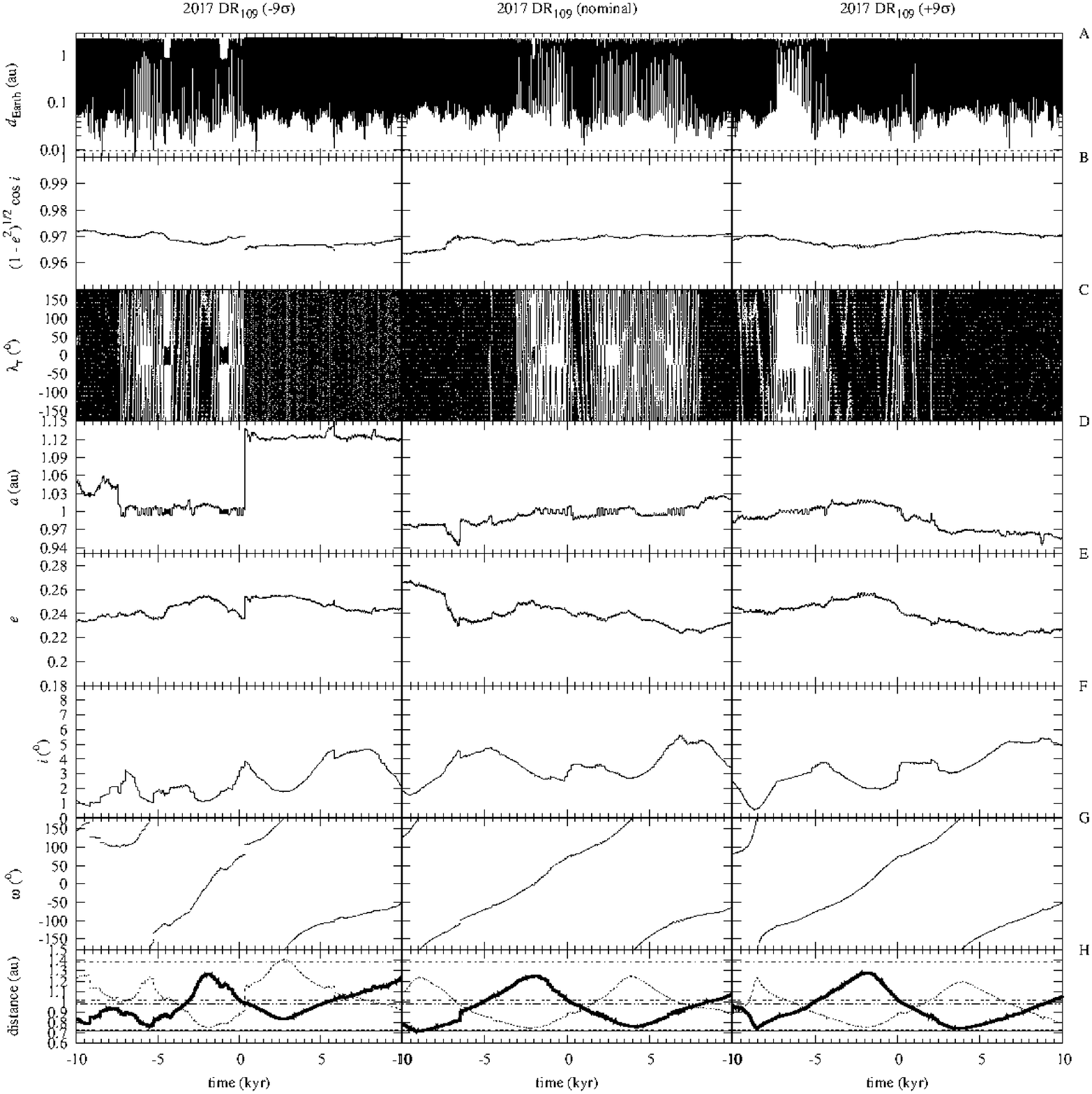}
        \caption{As Fig. \ref{control} but for 2017~DR$_{109}$.
                }
        \label{control2}
     \end{figure*}

        About 20 days before the discovery of 2017~FZ$_{2}$, another small NEA, 2017~DR$_{109}$, had been found following an orbit similar 
        to that of 2017~FZ$_{2}$.\footnote{http://www.minorplanetcenter.net/mpec/K17/K17E31.html} The new minor body was observed by D. C. 
        Fuls with the 0.68-m Schmidt camera of the Catalina Sky Survey at a visual apparent magnitude of 19.6 (Fuls et al. 2017). It is also 
        a possible co-orbital of our planet, $|a-a_{\rm Earth}|\sim$0.0011~au, but smaller (9--20~m) than 2017~FZ$_{2}$. Its current orbital 
        solution is not as robust as that of 2017~FZ$_{2}$ ---see Table~\ref{elements}, central column--- but it is good enough to arrive to 
        solid conclusions regarding its current dynamical status, in the neighbourhood of $t=0$. Fig.~\ref{orbitDR109} shows the behaviour 
        of the resonant angle of 2017~DR$_{109}$ during the time interval ($-$75, 75)~yr; it is indeed a co-orbital and follows a horseshoe 
        orbit with respect to our planet. Fig.~\ref{control2} indicates that the orbital evolution of this object is nearly as chaotic as 
        that of 2017~FZ$_{2}$. Although it may remain inside or in the neighbourhood of Earth's co-orbital zone for many thousands of years 
        (see D-panels), it switches between the various co-orbital states (quasi-satellite, Trojan or horseshoe) and hybrids of them 
        multiple times within the time interval displayed (see C-panels). Many of the dynamical aspects discussed regarding 2017~FZ$_{2}$ 
        are also present in Fig.~\ref{orbitDR109}. As for its next window of visibility, from 2018 February 23 to March 6 it will have an 
        apparent visual magnitude $<22$, moving from right ascension 0$^{\rm h}$ to 9$^{\rm h}$ and declination +60{\degr} to +10\degr, but 
        the Moon will interfere with the observations during this period. In general, objects following horseshoe-type paths with respect to 
        the Earth can be observed favourably only for a few consecutive years (often less than a decade), remaining at low solar elongations 
        ---and beyond reach of ground-based telescopes--- for many decades afterwards.
%
%
%
%
%
%
%
%

  \section{Are there too many of them?}
     The discovery of two small NEAs moving in rather similar orbits within 20 days of each other hints at the possible existence of a 
     dynamical grouping as the past and future orbital evolution of both objects also bears some resemblance. In order to search for 
     additional NEAs that may be following paths consistent with those of 2017~DR$_{109}$ and 2017~FZ$_{2}$, we use the D-criteria of 
     Southworth \& Hawkins (1963), $D_{\rm SH}$, Lindblad \& Southworth (1971), $D_{\rm LS}$ (in the form of equation 1 in Lindblad 1994 or 
     equation 1 in Foglia \& Masi 2004), Drummond (1981), $D_{\rm D}$, and the $D_{\rm R}$ from Valsecchi, Jopek \& Froeschl\'e (1999). 
     These criteria are customarily applied using proper orbital elements not osculating Keplerian orbital elements like those in Table 
     \ref{elements} (see e.g. Milani 1993, 1995; Milani \& Kne{\v z}evi{\'c} 1994; Kne{\v z}evi{\'c} \& Milani 2000; Milani et al. 2014, 
     2017). 

     Unfortunately, our exploration of the orbital evolution of 2017~DR$_{109}$ and 2017~FZ$_{2}$ in Section 2 indicates that, due to their 
     short Lyapunov times, it may not be possible to estimate meaningful proper orbital elements for objects moving in such chaotic orbits. 
     Proper elements are expected to behave as quasi-integrals of the motion, and are often computed as the mean semimajor axis, 
     eccentricity and inclination from a numerical simulation over a long timespan; when close planetary encounters are at work, no relevant 
     timespan can lead to parameters that remain basically unchanged. Although the dynamical context associated with these objects makes any 
     exploration of their possible past connections difficult, a representative set of orbits of a given pair with low values of the 
     D-criteria could be integrated to investigate whether their orbital evolution over a reasonable amount of time is also similar or not, 
     confirming or disproving any indirect indication given by the values of the D-criteria based on osculating Keplerian orbital elements. 
     Such an approach will be tested here. 
    
     \subsection{The evidence}
        Table~\ref{yorps}, top section, shows various orbital parameters and the values of the $D$-criteria for objects with $D_{\rm LS}$ 
        and $D_{\rm R} < 0.05$ with respect to 2017~FZ$_{2}$ as described by its nominal orbit in Table~\ref{elements}, left-hand column. 
        Unfortunately, the closest dynamical relatives of 2017~FZ$_{2}$ are also small NEAs with poor orbital determinations, 2009 HE$_{60}$
        and 2012 VZ$_{19}$. There are however in Table~\ref{yorps}, top section, some relatively large NEAs with well-constrained orbital 
        solutions, (488490) 2000~AF$_{205}$ and (54509)~YORP 2000~PH$_{5}$. Asteroid 488490 is considered as an accessible NEA suitable for 
        sample return missions (Christou 2003; Sears, Scheeres \& Binzel 2003). Besides YORP, 2017~DR$_{109}$ and 2017~FZ$_{2}$, two other 
        objects, 2009~HE$_{60}$ and 2015~YA, are also confined within Earth's co-orbital zone. Asteroid 2015~YA follows an asymmetric 
        horseshoe trajectory (a hybrid of quasi-satellite and horseshoe that may evolve into a pure quasi-satellite state or a plain 
        horseshoe one within the next century or so) with respect to the Earth, but it may not stay as Earth co-orbital companion for long 
        due to the very chaotic nature of its orbital evolution (de la Fuente Marcos \& de la Fuente Marcos 2016b); its orbit determination
        is as reliable as that of 2017~DR$_{109}$ and it is in need of improvement as well. 
%
%
     \begin{table*}
        \centering
        \fontsize{8}{11pt}\selectfont
        \tabcolsep 0.07truecm
        \caption{Orbital elements, orbital periods ($P$), perihelion ---$q = a \ (1 - e)$--- and aphelion ---$Q = a \ (1 + e)$--- distances, 
                 number of observations ($n$), data-arc span, absolute magnitudes ($H$) and MOID with the Earth of NEAs following orbits 
                 similar to that of 2017~FZ$_{2}$ (top section) and YORP (bottom section). The values of the various $D$-criteria 
                 ($D_{\rm SH}$, $D_{\rm LS}$, $D_{\rm D}$ and $D_{\rm R}$) are also displayed. The minor bodies are sorted by ascending 
                 $D_{\rm LS}$ and only those with $D_{\rm LS}$ and $D_{\rm R} < 0.05$ are shown. The orbits are referred to the epoch 2017 
                 September 4 as before, with the single exception of 2012 VZ$_{19}$ that is referred to the epoch 2456236.5 
                 (2012-November-05.0) TDB. Source: JPL's Small-Body Database.}
        \begin{tabular}{lllllllllllllllll}
           \hline
            Asteroid        & $a$ (au)  & $e$        & $i$ (\degr) & $\Omega$ (\degr) & $\omega$ (\degr) & $P$ (yr) & $q$ (au) & $Q$ (au)
                            & $n$ & arc (d) & $H$ (mag) & MOID (au)
                            & $D_{\rm SH}$ & $D_{\rm LS}$ & $D_{\rm D}$ & $D_{\rm R}$ \\
           \hline
            2009 HE$_{60}$  & 0.99581   & 0.26487    & 1.58478     & 229.00798        & 219.87710        &  0.99    & 0.7321   & 1.2596 
                            & 19  &  5      & 25.6      & 0.01544
                            & 0.5234       & 0.0100       & 0.2392      & 0.0209      \\
            2012 VZ$_{19}$  & 0.98250   & 0.25406    & 1.24739     &  57.95044        & 221.44994        &  0.97    & 0.7329   & 1.2321 
                            & 14  &  2      & 25.6      & 0.01111
                            & 0.0585       & 0.0163       & 0.0274      & 0.0435      \\
            488490          & 1.03408   & 0.27683    & 2.40837     & 220.06654        & 127.35081        &  1.05    & 0.7478   & 1.3204 
                            & 157 & 6339    & 21.7      & 0.01818
                            & 0.2768       & 0.0178       & 0.0954      & 0.0403      \\
            2017 BU         & 1.01959   & 0.25118    & 1.50522     &  75.57533        & 159.29112        &  1.03    & 0.7635   & 1.2757 
                            & 59  & 14      & 25.1      & 0.02293
                            & 0.2295       & 0.0263       & 0.0804      & 0.0339      \\
            2015 YA         & 0.99617   & 0.27938    & 1.61922     & 255.23303        &  83.61803        &  0.99    & 0.7179   & 1.2745 
                            & 50  & 5       & 27.4      & 0.00349
                            & 0.2449       & 0.0281       & 0.0865      & 0.0310      \\
            2008 UB$_{95}$  & 0.98970   & 0.26862    & 3.23715     &  21.96038        & 253.33450        &  0.98    & 0.7238   & 1.2556 
                            & 58  & 363     & 24.7      & 0.00034
                            & 0.1024       & 0.0307       & 0.0361      & 0.0364      \\
            2016 JA         & 1.01323   & 0.26845    & 0.05793     & 358.40711        & 112.58731        &  1.02    & 0.7412   & 1.2852 
                            & 46  & 10      & 27.6      & 0.00017
                            & 0.5331       & 0.0309       & 0.2590      & 0.0098      \\
            2017 DR$_{109}$ & 1.00064   & 0.24139    & 3.05996     & 341.31106        &  72.09400        &  1.00    & 0.7591   & 1.2422 
                            & 29  & 5       & 27.6      & 0.00619
                            & 0.4612       & 0.0362       & 0.1861      & 0.0199      \\
            2012 BD$_{14}$  & 1.01758   & 0.24045    & 1.53557     & 292.14513        & 293.67664        &  1.03    & 0.7729   & 1.2623 
                            & 61  & 10      & 26.5      & 0.00795
                            & 0.2610       & 0.0398       & 0.1001      & 0.0421      \\
            YORP            & 1.00622   & 0.23022    & 1.59931     & 278.28130        & 278.86596        &  1.01    & 0.7746   & 1.2379 
                            & 550 & 1826    & 22.7      & 0.00277
                            & 0.3526       & 0.0477       & 0.1426      & 0.0391      \\
            2000 QX$_{69}$  & 1.01012   & 0.27146    & 4.57402     & 150.13180        &  74.18700        &  1.02    & 0.7359   & 1.2843 
                            & 31  & 5       & 24.2      & 0.00020
                            & 0.2812       & 0.0491       & 0.0949      & 0.0175      \\
            2014 NZ$_{64}$  & 1.03678   & 0.27343    & 4.49454     & 162.44887        & 273.58758        &  1.06    & 0.7533   & 1.3203 
                            & 42  & 1046    & 22.8      & 0.00923
                            & 0.5218       & 0.0493       & 0.2250      & 0.0434      \\
           \hline
                            \multicolumn{17}{c}{with YORP}                            \\ 
           \hline
            2012 BD$_{14}$  & 1.01758   & 0.24045    & 1.53557     & 292.14513        & 293.67664        &  1.03    & 0.7729   & 1.2623 
                            & 61  & 10      & 26.5      & 0.00795
                            & 0.1172       & 0.0104       & 0.0434      & 0.0199      \\
            2017 BU         & 1.01959   & 0.25118    & 1.50522     &  75.57533        & 159.29112        &  1.03    & 0.7635   & 1.2757 
                            & 59  & 14      & 25.1      & 0.02293
                            & 0.1661       & 0.0238       & 0.0692      & 0.0255      \\
            2017 DR$_{109}$ & 1.00064   & 0.24139    & 3.05996     & 341.31106        &  72.09400        &  1.00    & 0.7591   & 1.2422 
                            & 29  & 5       & 27.6      & 0.00619
                            & 0.4512       & 0.0318       & 0.1907      & 0.0247      \\
            2010 FN         & 0.98958   & 0.21100    & 0.12365     & 161.55336        & 126.06353        &  0.98    & 0.7807   & 1.1984 
                            & 31  & 4       & 26.6      & 0.00082
                            & 0.3152       & 0.0327       & 0.1196      & 0.0350      \\
            2002 VX$_{91}$  & 0.98400   & 0.20127    & 2.34170     & 216.42578        &  78.74582        &  0.98    & 0.7859   & 1.1820 
                            & 45  & 1942    & 24.3      & 0.00180
                            & 0.3294       & 0.0337       & 0.1360      & 0.0481      \\
            471984          & 1.02619   & 0.22570    & 3.35738     &  49.72365        & 117.40597        &  1.04    & 0.7946   & 1.2578 
                            & 115 & 1020    & 22.7      & 0.02350
                            & 0.1441       & 0.0369       & 0.0488      & 0.0405      \\
            2007 WU$_{3}$   & 1.01129   & 0.20401    & 2.39332     & 177.68825        & 351.97450        &  1.02    & 0.8050   & 1.2176 
                            & 37  & 3164    & 23.8      & 0.04015
                            & 0.1232       & 0.0425       & 0.0736      & 0.0351      \\
            2009 BK$_{2}$   & 1.01251   & 0.21259    & 3.55344     & 126.08145        & 121.52628        &  1.02    & 0.7973   & 1.2278 
                            & 27  & 13      & 25.3      & 0.02604
                            & 0.2102       & 0.0446       & 0.0804      & 0.0227      \\
            2011 OJ$_{45}$  & 1.01688   & 0.20367    & 0.74918     & 288.97227        & 135.40425        &  1.03    & 0.8098   & 1.2240 
                            & 21  & 17      & 26.0      & 0.00755
                            & 0.4003       & 0.0465       & 0.1728      & 0.0500      \\
            2017~FZ$_{2}$   & 1.00714   & 0.26406    & 1.81167     & 185.86918        & 100.32304        &  1.01    & 0.7412   & 1.2731 
                            & 152 &  8      & 26.7      & 0.00139
                            & 0.3525       & 0.0477       & 0.1426      & 0.0391      \\
           \hline
        \end{tabular}
        \label{yorps}
     \end{table*}
%
%
%
%
     \begin{figure}
        \centering
        \includegraphics[width=\linewidth]{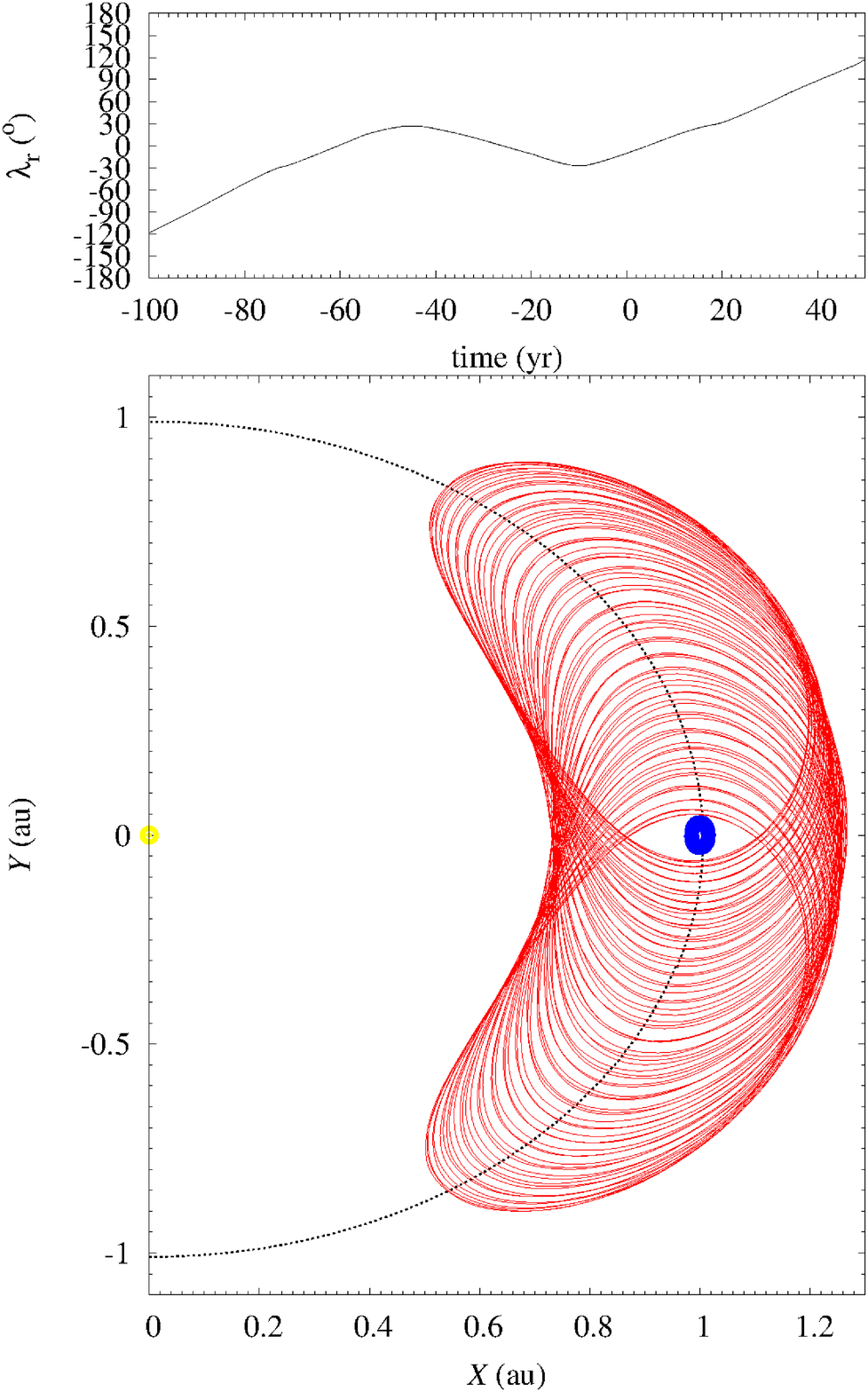}
        \caption{Similar to Fig. \ref{orbit} but for 2009~HE$_{60}$.
                }
        \label{orbitHE60}
     \end{figure}
%
%

        Although the quality of the orbital determination of 2009~HE$_{60}$ (Gibbs et al. 2009) is inferior to those of 2015~YA, 
        2017~DR$_{109}$ or 2017~FZ$_{2}$, our simulations show that this NEA is currently engaged in a brief quasi-satellite episode with 
        our planet that started nearly 70~yr ago and will end in about 15~yr from now, to become a regular horseshoe librator for over 
        10$^{3}$~yr. Fig. \ref{orbitHE60}, top panel, shows the evolution of the value of $\lambda_{\rm r}$; the bottom panel displays the 
        path followed by 2009~HE$_{60}$ in a frame of reference centred at the Sun and rotating with our planet, projected on to the 
        ecliptic plane. Fig. \ref{qsloops} shows ten loops of the analemmatic curve described by 2009~HE$_{60}$ (in green). We therefore 
        confirm that 2009~HE$_{60}$ is a robust candidate to being a current quasi-satellite of our planet. We speak of a candidate because 
        its orbital solution is in need of some improvement; unfortunately, it has not been observed since 2009 April 29. Its next window of 
        visibility spans from 2018 May 29 to June 9, when it will have an apparent visual magnitude $<23.5$, moving from right ascension 
        17$^{\rm h}$ to 16$^{\rm h}$ and declination $-$20{\degr} to $-$18\degr, but the Moon will interfere with the observations during 
        most of this period. In any case, our planet appears to be second to none regarding the number of known quasi-satellites (see the 
        review by de la Fuente Marcos \& de la Fuente Marcos 2016e).

        The presence of YORP among the list of candidates to being dynamical relatives of 2017~FZ$_{2}$ in Table~\ref{yorps}, top section, 
        makes one wonder whether some of those small NEAs may be the result of mass shedding from YORP itself or a putative larger object 
        that produced YORP. The data show that 2012 BD$_{14}$ appears to follow an orbit akin to that of YORP. Table~\ref{yorps}, bottom 
        section, explores the presence of NEAs moving in YORP-like orbits (the nominal orbit of YORP is given in Table~\ref{elements}, 
        right-hand column). The orbital similarity between YORP and 2012~BD$_{14}$ is confirmed; another relatively large NEA with good 
        orbital determination, (471984) 2013~UE$_{3}$, is also uncovered. The sample of known NEAs moving in YORP-like orbits comprises a 
        few objects with probable sizes larger than 100~m and several more down to the meteoroid size (a few metres). Some or all of them 
        could be trapped in a web of overlapping mean-motion and secular resonances as described by de la Fuente Marcos \& de la Fuente 
        Marcos (2016d), but the very chaotic nature of this type of orbits subjected to the recurrent direct perturbations of Venus and the 
        Earth--Moon system could favour an alternative scenario where some of these objects may have a past physical relationship. If some 
        of these NEAs had a common genetic origin, one may expect that their dynamical evolution back in time kept them within a relatively 
        small volume of the orbital parameter space.

\begin{figure}
        \centering
        \includegraphics[width=\linewidth]{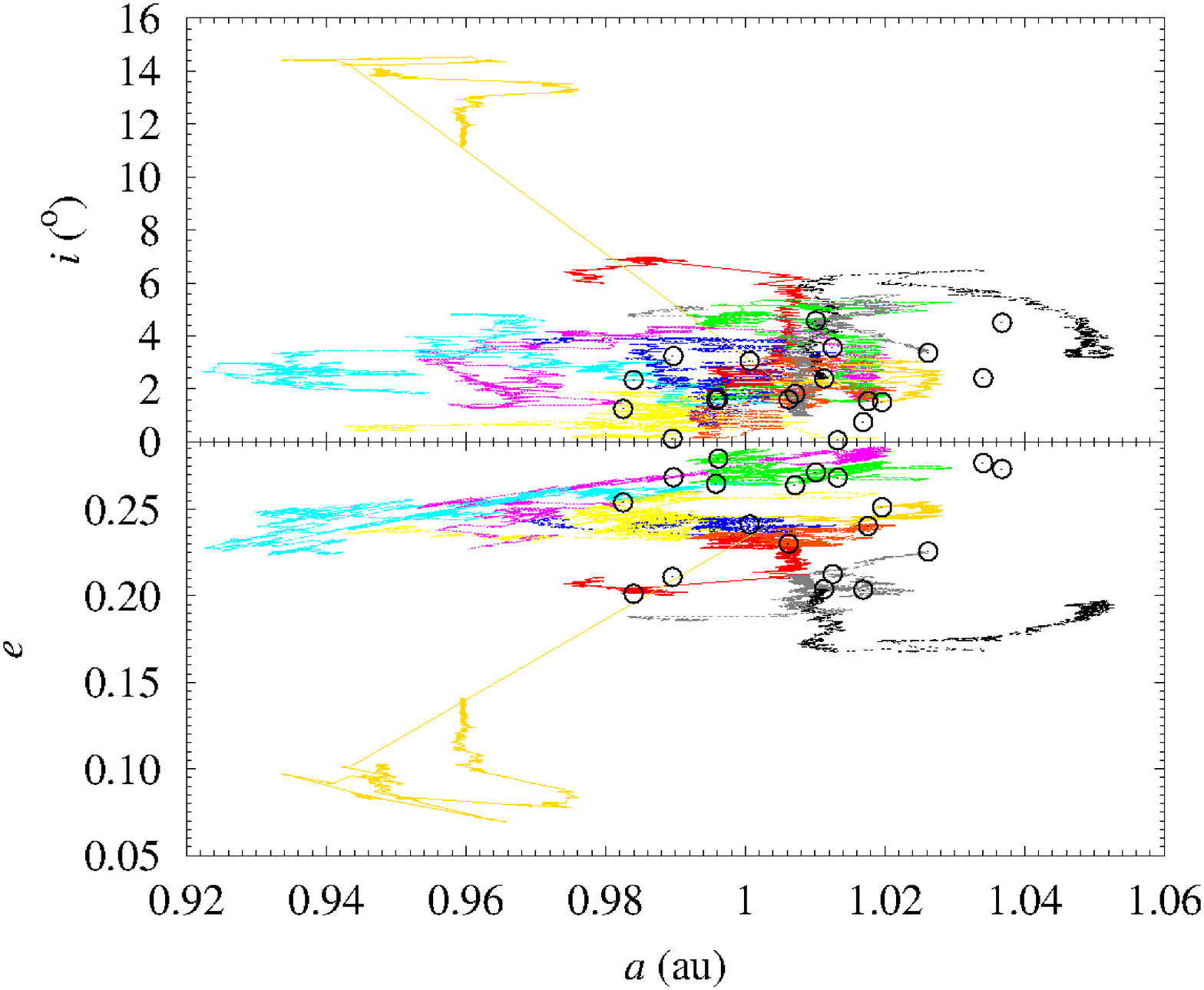}
        \caption{Backwards integrations of selected objects in Table \ref{yorps}. Black circles correspond to objects in Table \ref{yorps}
                 (19). The tracks are as follows: YORP (red), 2017~FZ$_{2}$ (green), 2017~DR$_{109}$ (blue), 2015~YA (pink), 2009~HE$_{60}$
                 (cyan), 2012~VZ$_{19}$ (yellow), 2007~WU$_{3}$ (black), 2012~BD$_{14}$ (orange), 471984 (grey) and 2017~BU (gold). These
                 integrations use initial conditions referred to to the epoch JD 2457800.5 (2017-February-16.0) TDB. 
                }
        \label{back}
     \end{figure}

        Fig. \ref{back} shows a number of backwards integrations for 10\,000 yr corresponding to the nominal orbits of representative NEAs
        in Table~\ref{yorps}. The points show the current parameters of the 19 different objects in Table~\ref{yorps}. As this figure uses
        only nominal orbits and some of the 19 NEAs have poor orbit determinations, the purpose of this plot is simply act as a guide to the
        reader, not to provide an extensive exploration of the backwards dynamical evolution of these objects, which is out of the scope of 
        this paper. In general, the backwards time evolution of these objects seems to keep them clustered within a relatively small region. 
        Some pairs follow unusually similar tracks, for example YORP and 2012~BD$_{14}$ or 471984 and 2007~WU$_{3}$. In principle, it is 
        difficult to conclude that the present-day orbits followed by some of these objects could be alike due to chance alone. But how 
        often do NEAs end up and remain on an orbit within this region of the orbital parameter space?
%
%
%
%

     \subsection{The expectations}
        In order to provide a statistically robust answer to the question asked in the previous section, we use the list of near-Earth 
        objects (NEOs) currently catalogued (as of 2017 July 20, 16\,498 objects, 16\,323 NEAs) by JPL's Solar System Dynamics Group (SSDG)
        Small-Body Database (SBDB),\footnote{http://ssd.jpl.nasa.gov/sbdb.cgi} concurrently with the NEOSSat-1.0 orbit model developed by 
        Greenstreet, Ngo \& Gladman (2012) and the newer one developed within the framework of the Near-Earth Object Population Observation 
        Program (NEOPOP) and described by Granvik et al. (2013a,b) and Bottke et al. (2014), which is intended to be a state-of-the-art 
        replacement for a widely used model described by Bottke et al. (2000, 2002). We use data from these two NEO models because such 
        synthetic data do not contain any genetically related objects and they are free from the observational biases and selection effects 
        that affect the actual data. These two features are critical in order to decide whether some of the NEAs in Table~\ref{yorps} may 
        have had a physical relationship in the past or not. 

\begin{figure*}
       \centering
        \includegraphics[width=0.33\linewidth]{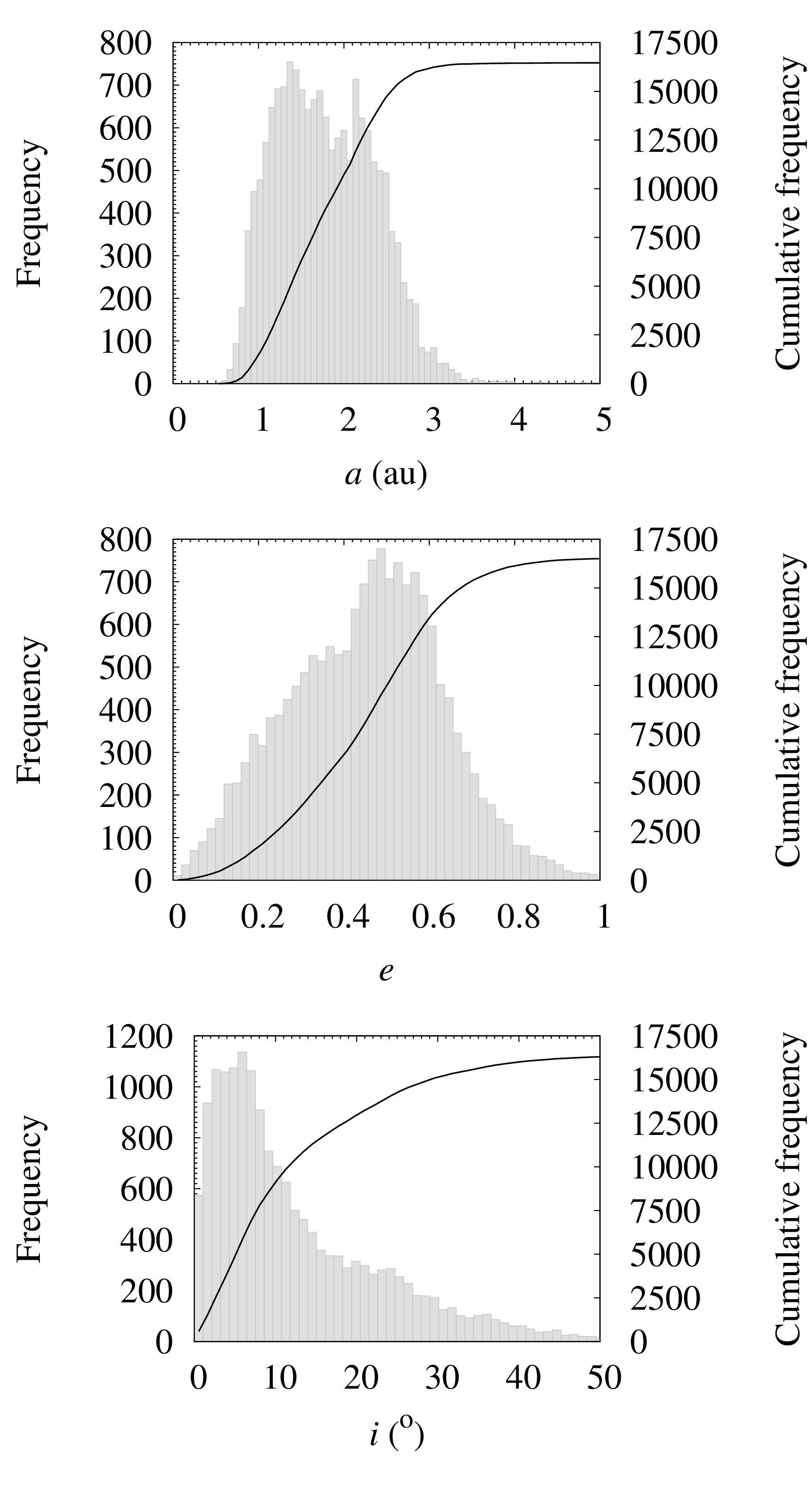}
        \includegraphics[width=0.33\linewidth]{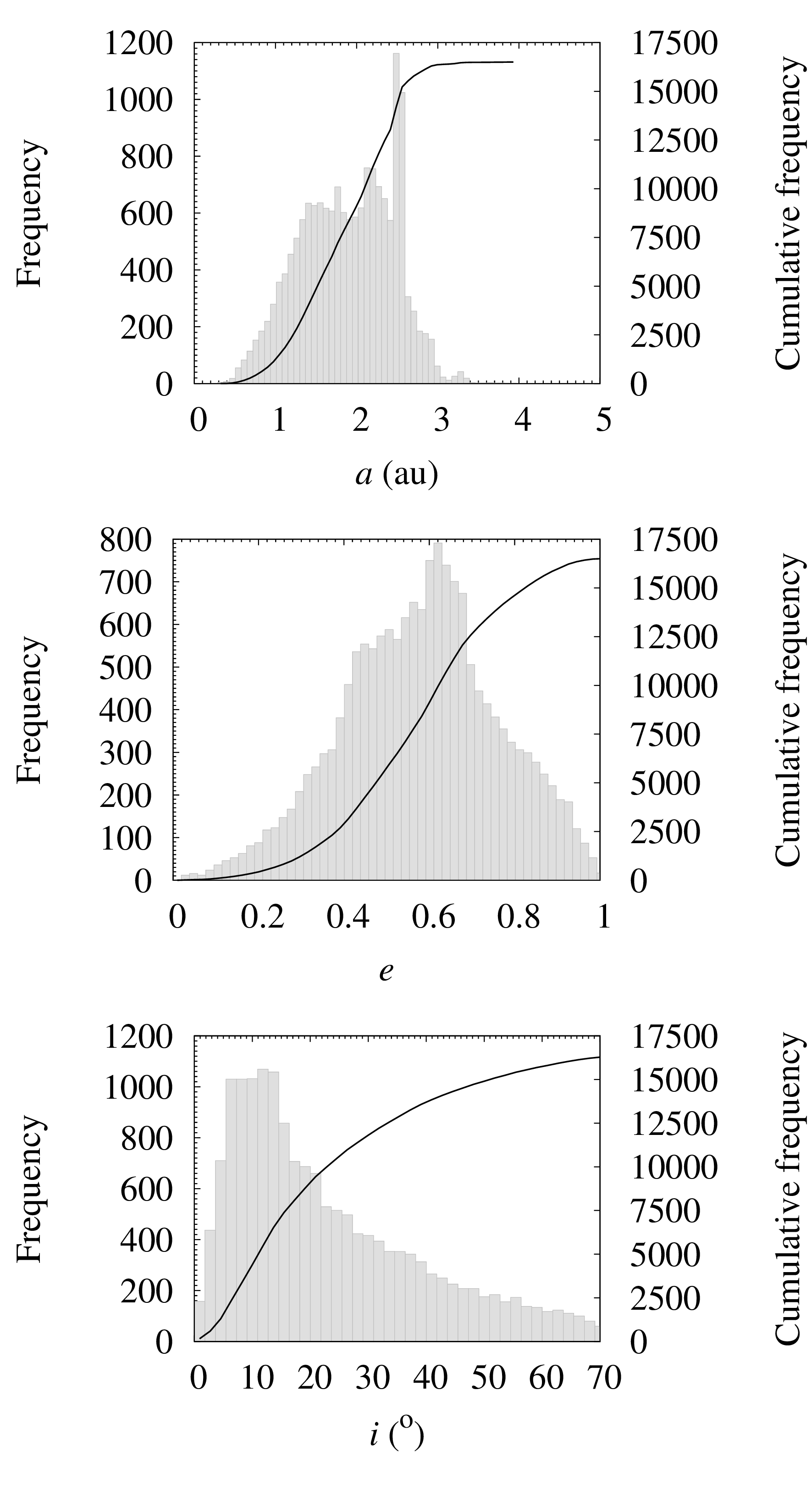}
        \includegraphics[width=0.33\linewidth]{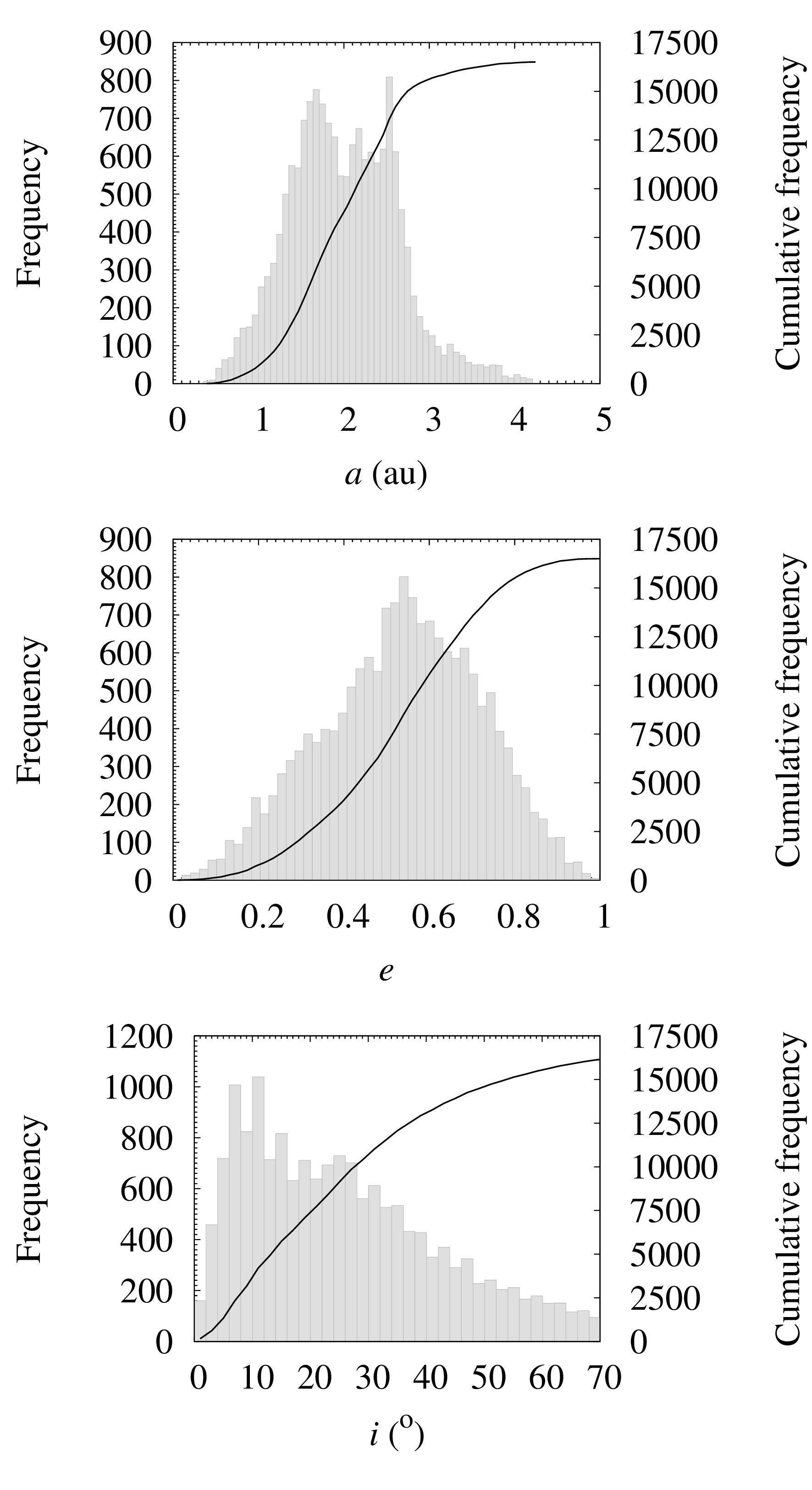}
        \caption{Distribution of the values of the orbital elements $a$, $e$, and $i$ for real NEOs (left-hand panels) and a representative 
                 synthetic population of NEOs from the NEOSSat-1.0 (central panels) and the NEOPOP (right-hand panels) models. The 
                 NEOSSat-1.0 model predicts an excess of minor bodies with values of the semimajor axis close to 2.5~au, which corresponds 
                 to objects leaving the main asteroid belt through the 3:1 mean-motion resonance with Jupiter (see the text for details). 
                 Data from JPL's SSDG SBDB (left-hand panels) as of 2017 July 20, 16\,498 NEOs.
                }
        \label{bias}
     \end{figure*}

        The data from JPL's SSDG SBDB (see Fig. \ref{bias}, left-hand panels)\footnote{In this and subsequent histograms, the bin size has 
        been computed using the Freedman-Diaconis rule (Freedman \& Diaconis 1981), i.e. $2\ {\rm IQR}\ n^{-1/3}$, where IQR is the 
        interquartile range and $n$ is the number of data points.} show that the probability of finding a NEA within Earth's co-orbital zone 
        is 0.0025$\pm$0.0004\footnote{The uncertainty has been computed assuming Poissonian statistics, $\sigma=\sqrt{n}$, see e.g. Wall \& 
        Jenkins (2012).} (40 objects out of 16\,323, but they may or may not be trapped in the 1:1 mean-motion resonance with our planet) 
        and that the probability of finding a NEA following a YORP-like orbit ---in other words, $D_{\rm LS}$ and $D_{\rm R}<0.05$ with 
        respect to YORP--- is 0.00067$\pm$0.00025\footnote{As before, we adopt Poissonian statistics to compute the uncertainty ---applying 
        the approximation given by Gehrels (1986) when $n<21$, $\sigma \sim 1 + \sqrt{0.75 + n}$.} (11 objects out of 16\,323). These are of 
        course biased numbers, but how do they compare with unbiased theoretical expectations?
%
%
%
%

        A quantitative answer to this question can be found looking at the predictions made by a scientifically validated NEO model. For 
        this task, we have used first the codes described in Greenstreet et al. (2012)\footnote{http://www.phas.ubc.ca/\%7Esarahg/n1model/} 
        with the same standard input parameters to generate sets of orbital elements including 16\,498 virtual objects. The NEOSSat-1.0 
        orbit model was originally applied to compute the number of NEOs with $H<18$~mag (or a diameter of about 1 km) in each dynamical 
        class ---Vatiras, Atiras, Atens, Apollos and Amors. It can be argued that only about 6.4 per cent of known NEOs have $H<18$~mag, but 
        if we assume that the size and orbital elements of asteroids are uncorrelated we can still use the NEOSSat-1.0 orbit model to 
        compute reliable theoretical probabilities ---this customary assumption has been contested by e.g. Bottke et al. (2014) and it is 
        not used by the NEOPOP model. Fig. \ref{bias}, central panels, shows a typical outcome from this NEO model. When we compare 
        left-hand and central panels in Fig. \ref{bias}, we observe that the two distributions are quite different in terms of the values of 
        the orbital inclination. There is also a significant excess of synthetic NEOs with values of the semimajor axis close to 2.5~au that 
        corresponds to minor bodies escaping the main asteroid belt through the 3:1 mean-motion resonance with Jupiter; the Alinda family of 
        asteroids are held by this resonance (see e.g. Simonenko, Sherbaum \& Kruchinenko 1979; Murray \& Fox 1984). 

        The NEOSSat-1.0 model predicts that the probability of finding a NEO within Earth's co-orbital zone is 0.0028$\pm$0.0003 and that of 
        a NEO following a YORP-like orbit is 0.00003$\pm$0.00003 (averages and standard deviations come from ten instances of the model 
        using different seeds to generate random numbers). It is clear that the values of the theoretical and empirical probabilities of 
        finding a NEO within Earth's co-orbital zone are statistically consistent; therefore, our assumption of lack of correlation between 
        size and orbital elements might be reasonably correct, at least within the context of this research. Surprisingly, the empirical 
        probability of finding a NEA following a YORP-like orbit is significantly higher than the theoretical one, over 21$\sigma$ higher. 
        While it can be argued that our approach is rather crude, it is difficult to explain such a large difference as an artefact when the 
        probabilities of finding a co-orbital agree so well (but see below for a more detailed discussion). 

        The NEOSSat-1.0 model is not the most recent and highest-fidelity NEO model currently available but, how different are the estimates 
        provided by a more up-to-date model like NEOPOP? The software implementing the NEOPOP model is publicly 
        available\footnote{http://neo.ssa.esa.int/neo-population} and the model has been successfully applied in several recent NEO studies 
        (Granvik et al. 2016, 2017). It accurately reproduces the orbits of the known NEOs as well as their absolute magnitudes and albedos, 
        i.e. size and orbital elements of NEOs are correlated in this framework. The model is calibrated from $H=15$~mag up to $H=25$~mag; 
        as our knowledge of the NEO population with $H<15$~mag is assumed to be complete, known NEOs have been used for this range of 
        absolute magnitudes. Fig.~\ref{bias}, right-hand panels, shows a typical outcome from this model. The synthetic, unbiased 
        distributions are quite different from the real ones (left-hand panels). The excess of synthetic NEOs with values of the semimajor 
        axis close to 2.5~au is less significant than that of the NEOSSat-1.0 model. Using the standard options of the NEOPOP model, the 
        code produces 802\,457 synthetic NEOs with $H<25$~mag. Out of this data pool, we select random permutations (only for $H>15$~mag, 
        the ones with $H<15$~mag remain unchanged) of 16\,498 objects to compute statistics or construct the histograms in Fig. \ref{bias}, 
        right-hand panels.

        The NEOPOP model predicts that the probability of finding a NEO within Earth's co-orbital zone is 0.0021$\pm$0.0004 and that of a
        NEO following a YORP-like orbit is 0.000024$_{-0.000024}^{+0.000043}$ (averages and standard deviations come from 25 instances of 
        the model). These two values of the probability agree well with those derived from the NEOSSat-1.0 model. Therefore, the theoretical 
        and empirical probabilities of finding a NEO within Earth's co-orbital zone are statistically consistent, but the empirical one of 
        finding a NEA following a YORP-like orbit is well above any of the theoretical ones. It could be argued that the theoretical and 
        empirical probabilities of finding a co-orbital are equal by chance. In addition, known NEAs in YORP-like orbits are simply more 
        numerous in relative terms because they tend to pass closer to our planet and in consequence are easier to discover as accidental 
        targets of NEO surveys. Although these arguments have significant weight, one can also argue that all the known Earth co-orbitals 
        are serendipitous discoveries that were found as a result of having experienced very close encounters with our planet.

        Regarding the issue of observed and predicted absolute magnitudes, let us focus only on NEOs, real or synthetic, with $H<25$~mag. 
        The probabilities given by the NEOPOP model are the values already cited. The data from JPL's SSDG SBDB give a value of 
        0.0021$\pm$0.0004 (27 out of 12736 objects) for the probability of finding a NEA with $H<25$~mag within Earth's co-orbital zone, 
        which is a perfect match for the theoretical estimate from the NEOPOP model. As for YORP-like orbits with $H<25$~mag, the empirical
        value of the probability is 0.0003$_{-0.0003}^{+0.0002}$ (4 out of 12736 objects). Even if we take into account the values of $H$,
        the theoretical value of the probability of observing a YORP-like orbit is significantly below the empirical value. In addition, the
        NEOPOP model predicts no NEOs with $H<23.5$~mag moving in YORP-like orbits but there are two known, YORP and 471984. It is true that 
        the evidence is based on small samples but one tantalizing albeit somewhat speculative possibility is that some of these objects
        have been produced in situ via fragmentation events, i.e. they are direct descendants of minor bodies that were already moving in 
        YORP-like orbits.

        On the other hand, the synthetic data generated to estimate the theoretical values of the probabilities can be used together with 
        the D-criteria to uncover unusual pairs among the NEAs in Table~\ref{yorps}. The procedure is simple, each set of synthetic NEOs 
        from NEOSSat-1.0 or NEOPOP comprises fully unrelated virtual objects. If we compute the D-criteria with respect to the real YORP (or 
        2017~FZ$_{2}$) for all the virtual NEOs in a given set and perform a search looking for those with $D_{\rm LS}$ and $D_{\rm R}<0.05$, 
        we can extract a sample of relevant unrelated (synthetic) NEAs and compute the expected average values and standard deviations of 
        the D-criteria when no orbital correlation between pairs of objects is present in the data. These averages and dispersions can be 
        used to estimate how unlikely (in orbital terms) the presence of a given pair of real objects is. 

        In order to evaluate statistically this likelihood, we compute the absolute value of the difference between the value of a certain 
        D-criterion for a given pair of objects in Table~\ref{yorps} and the average value for uncorrelated virtual objects, then divide by 
        the dispersion found from the synthetic data. If the estimator gives a value close to or higher than 2$\sigma$ we may assume that 
        the pair of objects could be unusual (an outlier, in statistical parlance) within the fully unrelated pairs scenario and perform 
        additional analyses. We have carried out this investigation and the estimator discussed gives values around 1$\sigma$ for all the 
        tested pairs with one single exception, the pair YORP--2012~BD$_{14}$ that gives values of the estimator in the range 2--3$\sigma$ 
        for all the D-criteria.

        A simple scenario that may explain the observed excess or even the presence of the unusual pair YORP--2012~BD$_{14}$ is in-situ 
        production of NEAs moving in YORP-like orbits either as a result of mass shedding from YORP itself (or any other relatively large 
        NEA in this group) or a putative bigger object that produced YORP (or any other of the larger objects in Table~\ref{yorps}). 
        However, it is not just the size of this population that matters: we also have the issue of its stability and how many concurrent 
        resonant objects should be observed in these orbits at a given time. Finding answers to these questions requires the analysis of 
        relevant $N$-body simulations. Given the chaotic nature of these orbits, relatively short calculations should suffice to arrive to 
        robust conclusions.

\begin{figure}
        \centering
        \includegraphics[width=\linewidth]{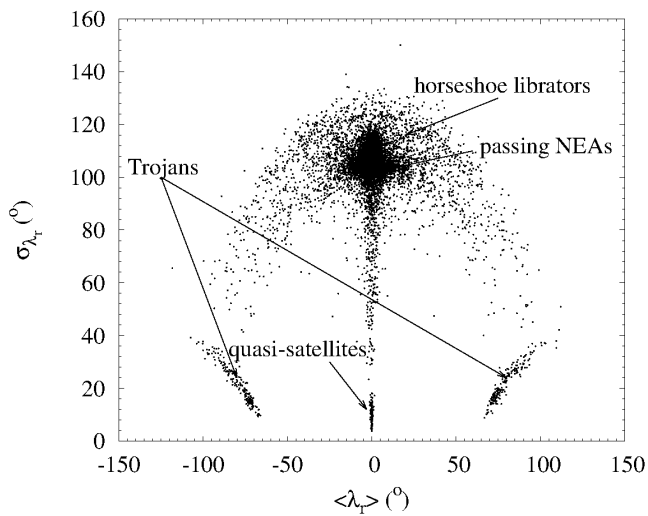}
        \caption{Standard deviation of the resonant angle as a function of the average value of the resonant angle for a set of 20\,000
                 numerical experiments involving virtual NEAs moving in YORP-like orbits (see the text for details). The main resonant
                 and non-resonant dynamical classes are indicated. 
                }
        \label{all}
     \end{figure}

\begin{figure}
        \centering
        \includegraphics[width=\linewidth]{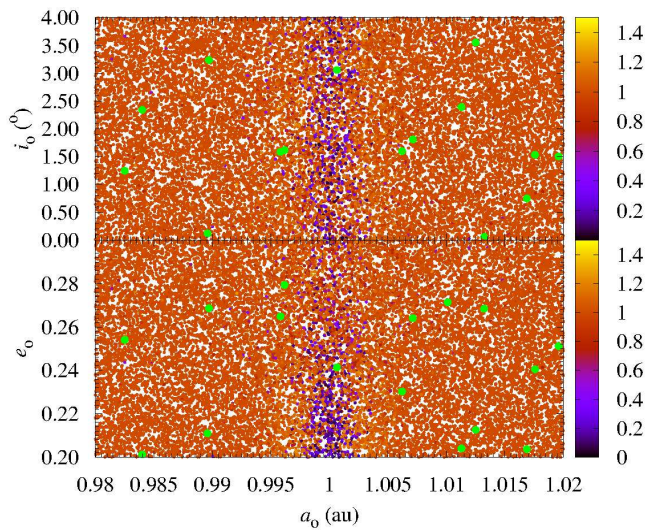}
        \caption{Resonant behaviour of the sample of virtual NEAs in Fig. \ref{all} in terms of the initial values of $a$, $e$ and $i$. The 
                 colours in the colour maps are proportional to the ratio between the standard deviation of the resonant angle and the 
                 expected value for a non-resonant behaviour (103\fdg923); green points correspond to objects in Table \ref{yorps}.
                }
        \label{coorb}
     \end{figure}

        In order to explore the stability and resonant properties of YORP-like orbits, we have performed 20\,000 numerical experiments using 
        the same software and physical model considered in the previous section and orbits for the virtual NEOs uniformly distributed (i.e 
        we do not make any assumptions on the origin, natural versus artificial, of the virtual objects) within the ranges 
        $a\in(0.98, 1.02)$~au, $e\in(0.20, 0.30)$, $i\in(0, 4)$\degr, $\Omega\in(0, 360)$\degr, and $\omega\in(0, 360)$\degr, i.e. 
        consistent with having $D_{\rm LS}$ and $D_{\rm R}<0.05$ with respect to YORP. After an integration of 2\,000 yr forward in time we 
        obtain Figs \ref{all} and \ref{coorb}. For each simulated particle we compute the average value and the standard deviation of the 
        resonant angle in the usual way (see e.g. Wall \& Jenkins 2012); the standard deviation of a non-resonant angle is 103\fdg923. In 
        Fig. \ref{all}, nearly 66 per cent of the virtual NEOs do not exhibit any sign of resonant behaviour during the simulation, 0.7 per 
        cent are quasi-satellites, nearly 1.8 per cent are Trojans, and 5.6 per cent move in horseshoe orbits, the remaining 25.9 per cent 
        follow elementary or compound 1:1 resonant states for at least some fraction of the simulated time. Fig. \ref{coorb} shows that the 
        most stable configurations are invariably characterized by very long synodic periods (i.e. $a$$\sim$1~au) and lower $e$. 

        From Table~\ref{yorps}, bottom section, and results in Sections 2 and 3, as well as results from the literature, we have at least 
        three objects held by the 1:1 mean-motion resonance with our planet out of 11 listed NEAs. In other words, over 27 per cent of all 
        the objects moving in YORP-like orbits exhibit signs of resonant behaviour when the theoretical expectation in the framework of a 
        fully random scenario is about 34 per cent. This result can be regarded as another piece of evidence in support of our previous 
        interpretation because the orbital distribution of NEAs is not expected to be random (in the sense of uniform), see Fig.~\ref{bias}, 
        and real NEAs are not supposed to have formed in these peculiar orbits, but come from the main asteroid belt. In addition we have 
        found one quasi-satellite (2017~FZ$_{2}$) out of 11 NEAs in Table~\ref{yorps}, bottom section, and two NEAs moving in horseshoe-type
        paths (YORP and 2017~DR$_{109}$), i.e. a quasi-satellite probability of about 9 per cent when the random scenario predicts 0.7 per 
        cent and a horseshoe probability of about 18 per cent when the random scenario predicts 5.6 per cent.
%
%
%
%
%
%
%
%

        The various aspects discussed above strongly suggest that some of the NEAs in Table~\ref{yorps} may form a dynamical grouping and
        perhaps have a genetic connection. Evidence for the presence of possible dynamical groupings among the co-orbital populations of our 
        planet has been discussed in the case of NEAs moving in Earth-like orbits, the Arjunas (de la Fuente Marcos \& de la Fuente Marcos 
        2015a), and also the group of objects following paths similar to those of 2015 SO$_{2}$, (469219) 2016~HO$_{3}$ and 2016 CO$_{246}$ 
        (de la Fuente Marcos \& de la Fuente Marcos 2016a,f). However, the study by Schunov{\'a} et al. (2012) could not find any
        statistically significant group of dynamically related NEAs among those known at the time, but it is also true that many objects in
        Table~\ref{yorps} have been discovered after the completion of that study.  

  \section{Mutual close encounters: happening by chance or not?}
     If the excess of NEAs following YORP-like orbits is statistically significant, one would expect that some of the asteroid pairs under
     study had their orbits intersecting one another in the past and perhaps that their relative velocities were very low during such very 
     close approaches. An exploration of this scenario requires the analysis of large sets of $N$-body simulations backwards in time for 
     each relevant pair, but in order to identify any results as statistically significant we must study first what happens in the case of 
     a random population of virtual NEAs moving in YORP-like orbits. The statistical analysis of minimum approach distances and relative 
     velocities between unrelated objects can help us distinguish between close encounters happening by chance and those resulting from an 
     underlying dynamical (and perhaps physical) relationship. 

     Using a sample of 10\,000 pairs of unrelated, virtual NEAs with orbits uniformly distributed within the ranges considered in the 
     numerical experiments performed in the previous section, and integrated backwards in time for 2\,000~yr, we have studied how is the 
     distribution of relative velocities at minimum approach distance ($<0.1$~au, the first quartile is 0.0046~au and the average is 
     0.017~au) during mutual encounters. Given the fact that these NEAs are too small to change significantly their orbits due to their 
     mutual gravitational attraction during close encounters, such a velocity distribution can be considered as a reasonably robust proxy 
     for the representative collision velocity distribution for pairs of NEAs following YORP-like orbits. The results of this analysis may 
     apply to both natural and artificial objects.
%
%
      \begin{figure*}
        \centering
         \includegraphics[width=0.33\linewidth]{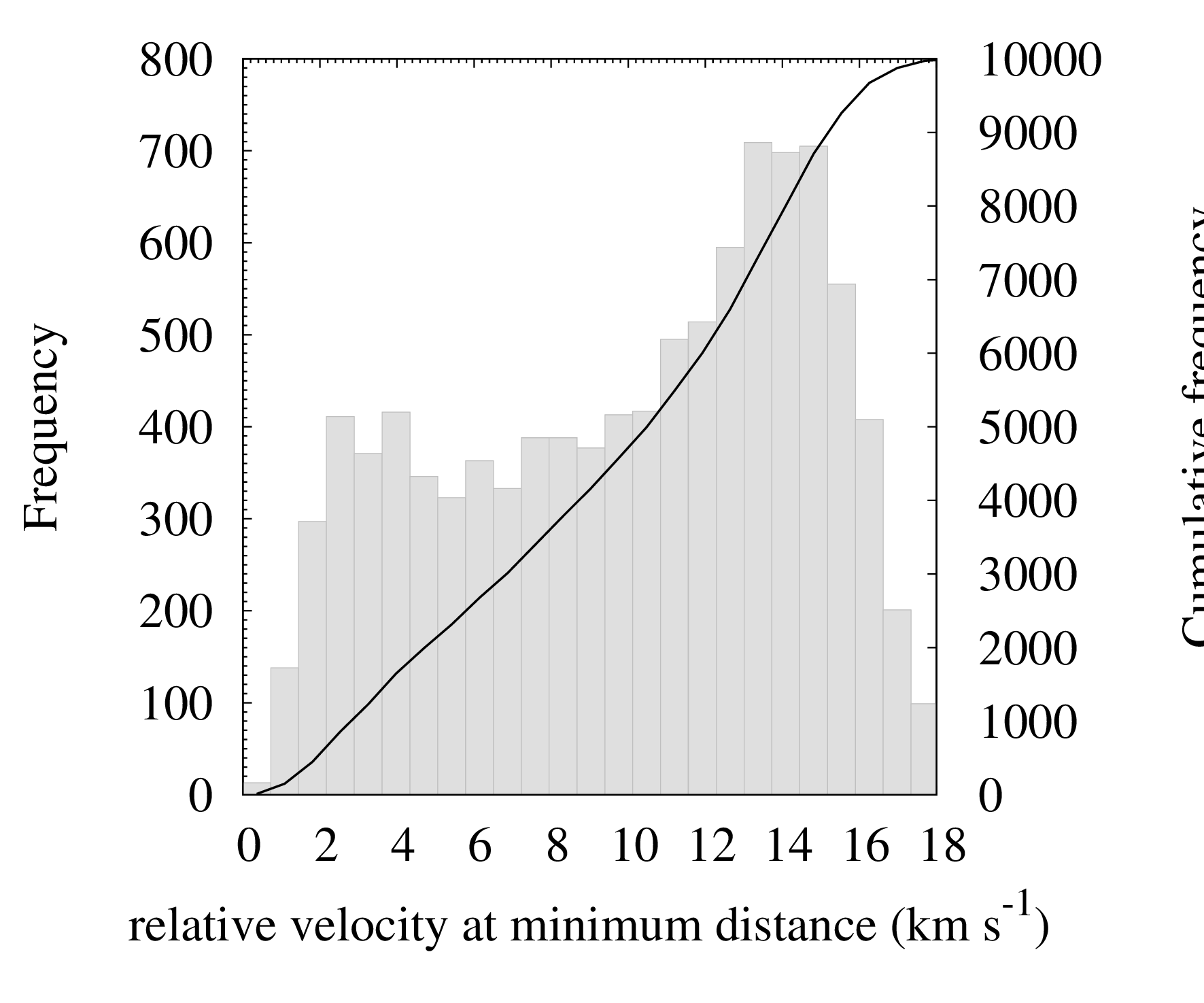}
         \includegraphics[width=0.33\linewidth]{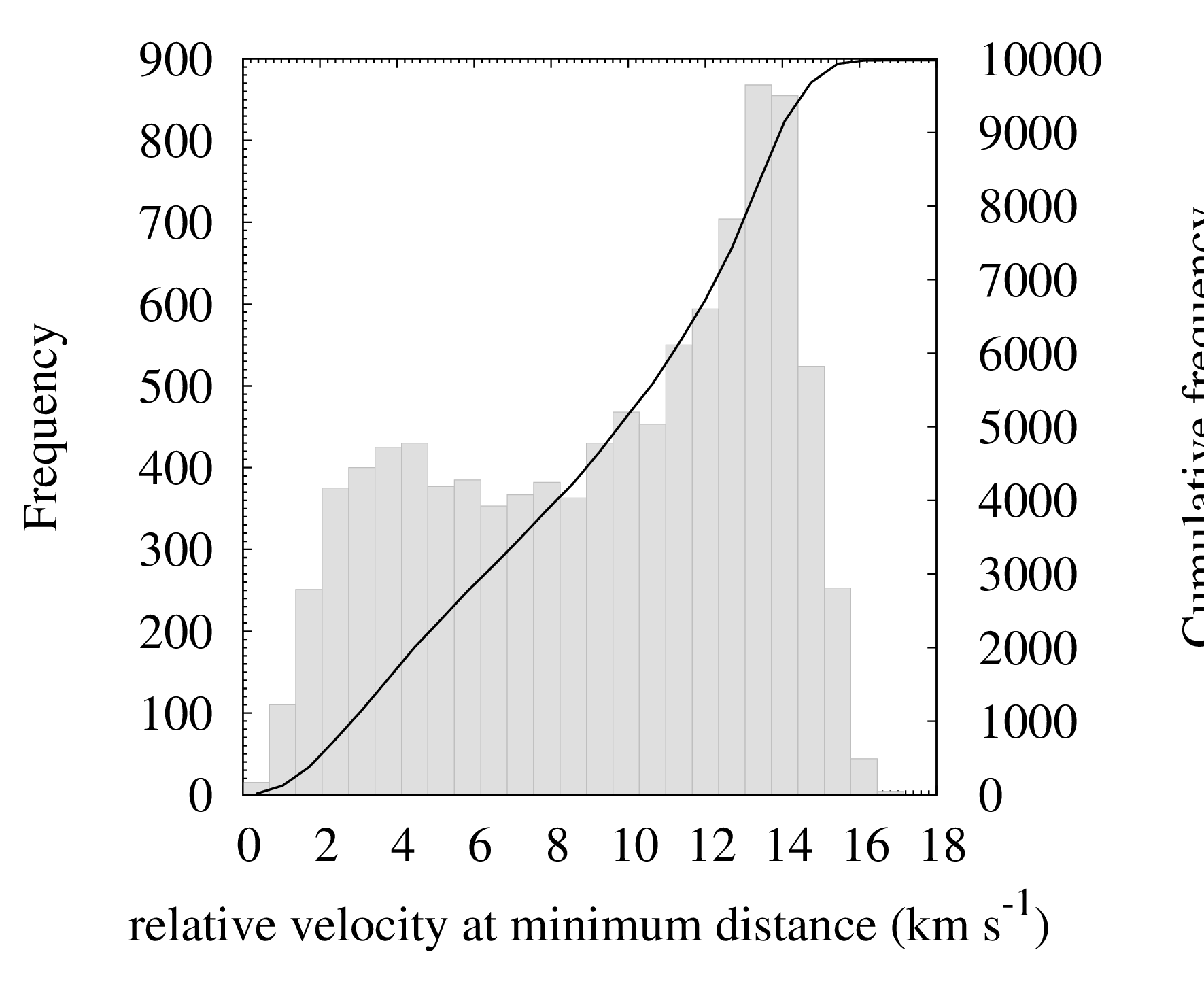}
         \includegraphics[width=0.33\linewidth]{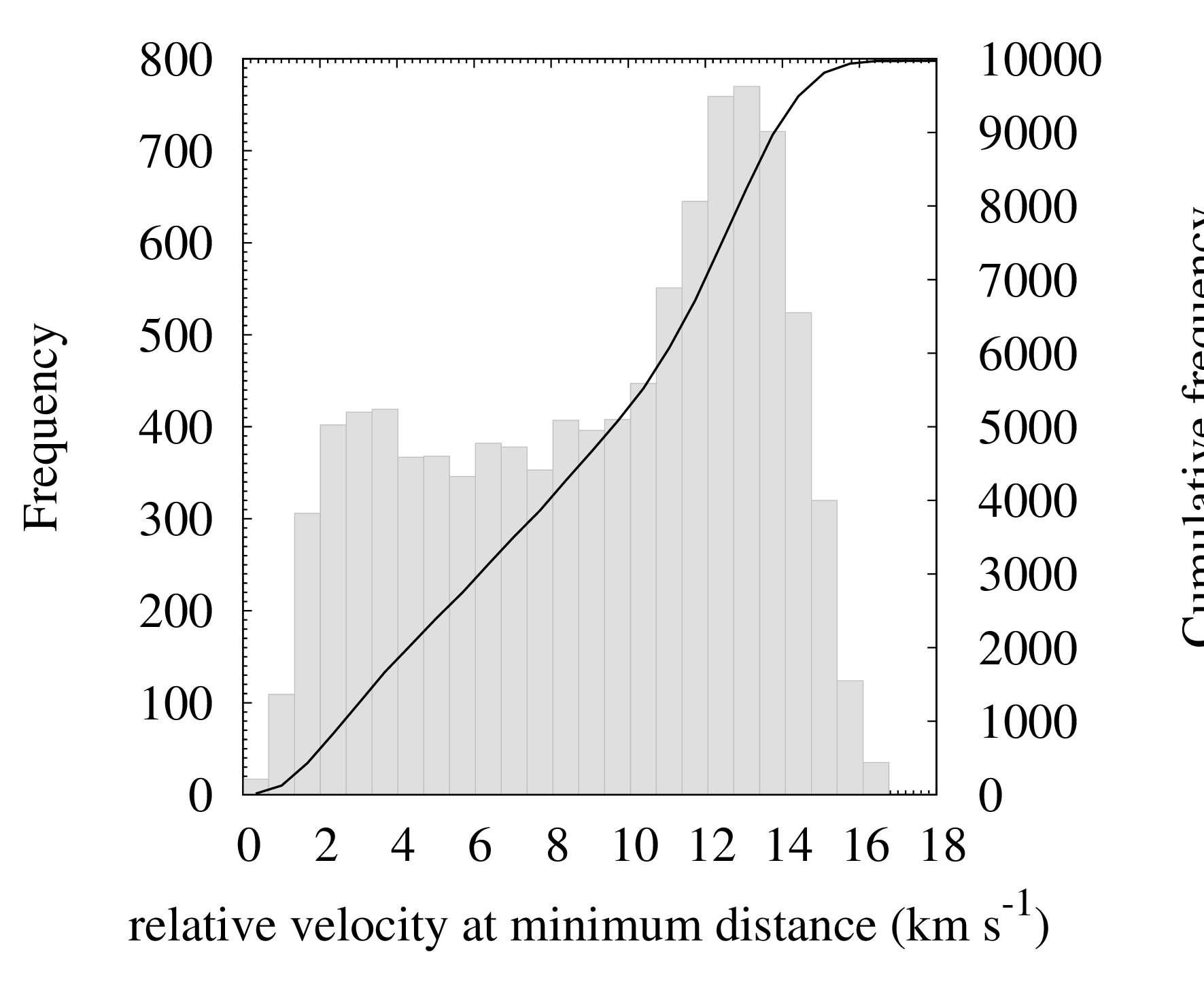}
         \caption{Velocity distribution between two virtual NEAs following YORP-like orbits (see the text for details). Orbits uniformly 
                  distributed as in Section 3.2 (left-hand panels); in this case, the bin width is 0.72259~km~s$^{-1}$ with IQR = 
                  7.78388~km~s$^{-1}$. Orbits from the NEOPOP model (central panels); bin width is 0.68598~km~s$^{-1}$ with IQR =
                  7.38954~km~s$^{-1}$. Orbits uniformly distributed but with ranges matching those of the synthetic sample from the NEOPOP 
                  model (right-hand panels); bin width is 0.67060~km~s$^{-1}$ with IQR = 7.22377~km~s$^{-1}$.
                 }
         \label{rvdist}
      \end{figure*}
%
%

     Fig.~\ref{rvdist}, left-hand panels, shows the distribution of relative velocities at minimum distance for pairs of virtual NEAs in 
     YORP-like orbits like those in the numerical experiments performed in the previous section. The most likely encounter velocities are 
     near the high velocity extreme, which is characteristic of encounters near perihelion where multiple encounter geometries are possible. 
     The mean approach velocity falls in the low-probability dip between the high-probability peak at about 14~km~s$^{-1}$ (encounters at 
     perihelion) and the secondary peak at about 3~km~s$^{-1}$ (encounters at aphelion). However, the closest approaches (at about 
     10\,000~km from each other) are characterized by values of the relative velocity close to 13~km~s$^{-1}$ due to their relatively high 
     eccentricity. The results in Fig.~\ref{rvdist}, left-hand panels, do not take into account predictions from the NEOPOP model and this
     may be seen as a weakness of our analysis. In order to explore possible systematic differences, we have completed an additional set of 
     calculations using the actual synthetic orbits predicted by the NEOPOP model as input, but randomizing the values of $\Omega$ and 
     $\omega$, then picking up pairs at random to compute the new set of simulations. The results of these calculations are shown in 
     Fig.~\ref{rvdist}, central panels, and are quite similar to those in the left-hand panels. However, the ranges in the values of $a$, 
     $e$ and $i$ are slightly different. We have repeated the calculations using uniform orbital distributions, but for the ranges 
     $a\in(0.985, 1.035)$~au, $e\in(0.19, 0.27)$, and $i\in(0, 4.5)$\degr; Fig.~\ref{rvdist}, right-hand panels, shows that the results are 
     very much alike. Although no actual collisions have been observed during all these simulations, the velocity distribution in 
     Fig.~\ref{rvdist} is the one we seek as the mutual gravitational perturbation is negligible even during the closest mutual flybys. 

     Fig.~\ref{rvdist} shows that the mean value of the velocity and its dispersion do not represent the distribution well. Collisions 
     between these minor bodies yield higher probable collision velocities although high speed lowers the overall collision probability (the 
     probability of experiencing an encounter closer than 30\,000~km is of the order of 0.0002). However, impacts from meteoroids originated 
     within this population on members of the same population happen at most probable velocities high enough to perhaps induce catastrophic 
     disruption or at least partial destruction of the target as the impact kinetic energy goes as the square of the collision velocity. Our 
     results closely resemble those obtained by Bottke et al. (1994) for the velocity distributions of colliding asteroids in the main 
     asteroid belt.

     YORP-induced catastrophic asteroid breakups can generate multiple small NEAs characterized by pair-wise velocity dispersions under 
     1~m~s$^{-1}$ and also clouds of dust as observed in the case of P/2013 R3 (Jewitt et al. 2017). Groups of objects moving initially 
     along similar paths lose all dynamical coherence in a rather short time-scale (Pauls \& Gladman 2005; Rubin \& Matson 2008; Lai et al. 
     2014). This randomization is accelerated when recurrent close planetary encounters are possible. As the virtual NEAs used in our 
     numerical experiments move in uncorrelated orbits, the results in Fig.~\ref{rvdist} can be used to single out pairs of objects that 
     exhibit some level of orbital coherence; this feature may signal a very recent (not more than a few thousand years ago) breakup event. 
     The velocity distribution of any pair of unrelated NEAs following YORP-like orbits is expected to be strongly peaked towards the high 
     end of the distribution in Fig.~\ref{rvdist}, i.e. yield a most probable collision velocity close to or higher than 13~km~s$^{-1}$. We 
     have explored the velocity distributions of multiple pairs of objects in Table \ref{yorps} using the same approach applied before and 
     found that this is true for the vast majority of them. However, we have found two notable exceptions, the pairs (54509)~YORP 
     2000~PH$_{5}$--2012~BD$_{14}$ and YORP--2007~WU$_{3}$. The first one has a non-negligible probability of experiencing close encounters 
     with relative velocities well under 1~km~s$^{-1}$, 0.0007. The second one has a most probable close approach velocity $<4$~km~s$^{-1}$
     ---the probability of a value of the close approach velocity that low or lower is 0.77---  well below the high-probability peak at 
     14~km~s$^{-1}$ in Fig.~\ref{rvdist}. The probability of having a close approach velocity $<4$~km~s$^{-1}$ from Fig.~\ref{rvdist} is 
     0.14.

\begin{figure}
        \centering
         \includegraphics[width=\linewidth]{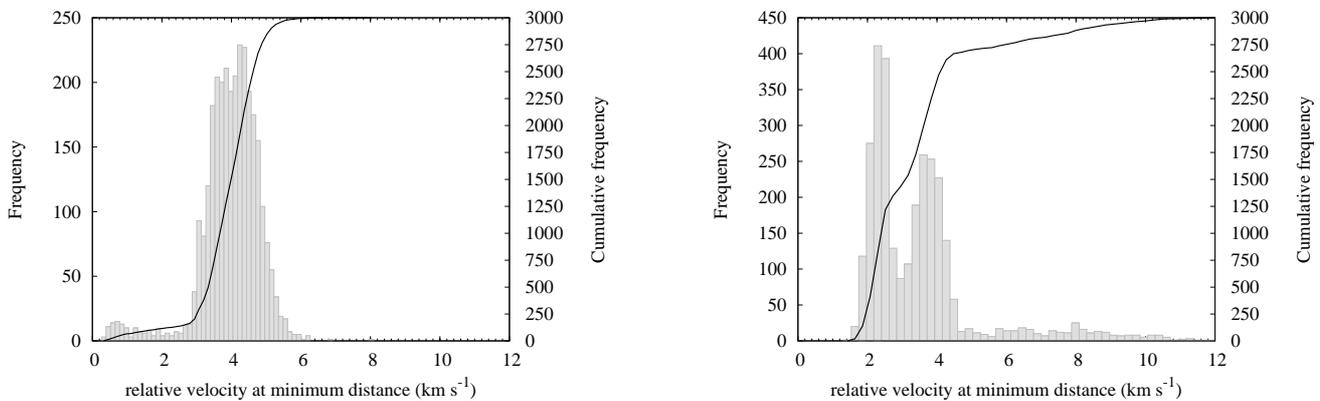}
         \caption{Velocity distribution between two virtual NEAs following control orbits statistically compatible with those of 
                  (54509)~YORP 2000~PH$_{5}$ and 2012~BD$_{14}$. Here, the bin width is 0.1306~km~s$^{-1}$, with IQR = 0.9416~km~s$^{-1}$ 
                  (see the text for details). 
                 }
         \label{YB}
      \end{figure}

     Asteroid 2012~BD$_{14}$ has a relatively good orbit (Holmes et al. 2012) as it has one radar Doppler observation, but lacks any 
     additional data other than those derived from the available astrometry/photometry. Assuming similar composition, this Apollo asteroid 
     must be as large as 2017~FZ$_{2}$ (15--33~m). In order to compute the velocity distribution for the pair YORP--2012~BD$_{14}$, we have 
     performed an analysis analogous to that in Fig.~\ref{rvdist} but using control orbits for both objects generated by applying the MCCM 
     method as described above and integrating backwards for 1\,000~yr. Fig.~\ref{YB} shows that the most probable collision velocity in 
     this case is $\sim$4~km~s$^{-1}$ and about 7 per cent of flybys have relative velocities close to or below 2~km~s$^{-1}$, but down to 
     336~m~s$^{-1}$. In addition, the probability of experiencing an encounter closer than 30\,000~km is 0.0013, which is significantly 
     higher than that of analogous encounters for random YORP-like orbits. In other words, it is possible to find control orbits ---albeit 
     with relatively low probability--- of YORP and 2012~BD$_{14}$, statistically compatible with the available observations, that place 
     them very close to each other and moving at unusually low relative speed in the recent past. These approaches are observed at 
     0.75--0.9~au from the Sun, i.e. in the inmediate neighbourhood of Venus. Although outside the scope of this paper, it may be possible 
     to find control orbits of these two objects that may lead to grazing encounters at velocities of a few m~s$^{-1}$ if a much larger 
     sample of orbits is analysed (the current one consists of 3\,000 virtual pairs). It is difficult to explain such a level of coherence 
     in their recent past orbital evolution as due to chance causes; these two objects could be genetically related. Unfortunately, 
     2012~BD$_{14}$ reached its most recent visibility window late in summer 2017, from August 25 to September 18, but this NEA will become 
     virtually unobservable from the ground during the next few decades.
%
%
%
%

\begin{figure}
        \centering
         \includegraphics[width=\linewidth]{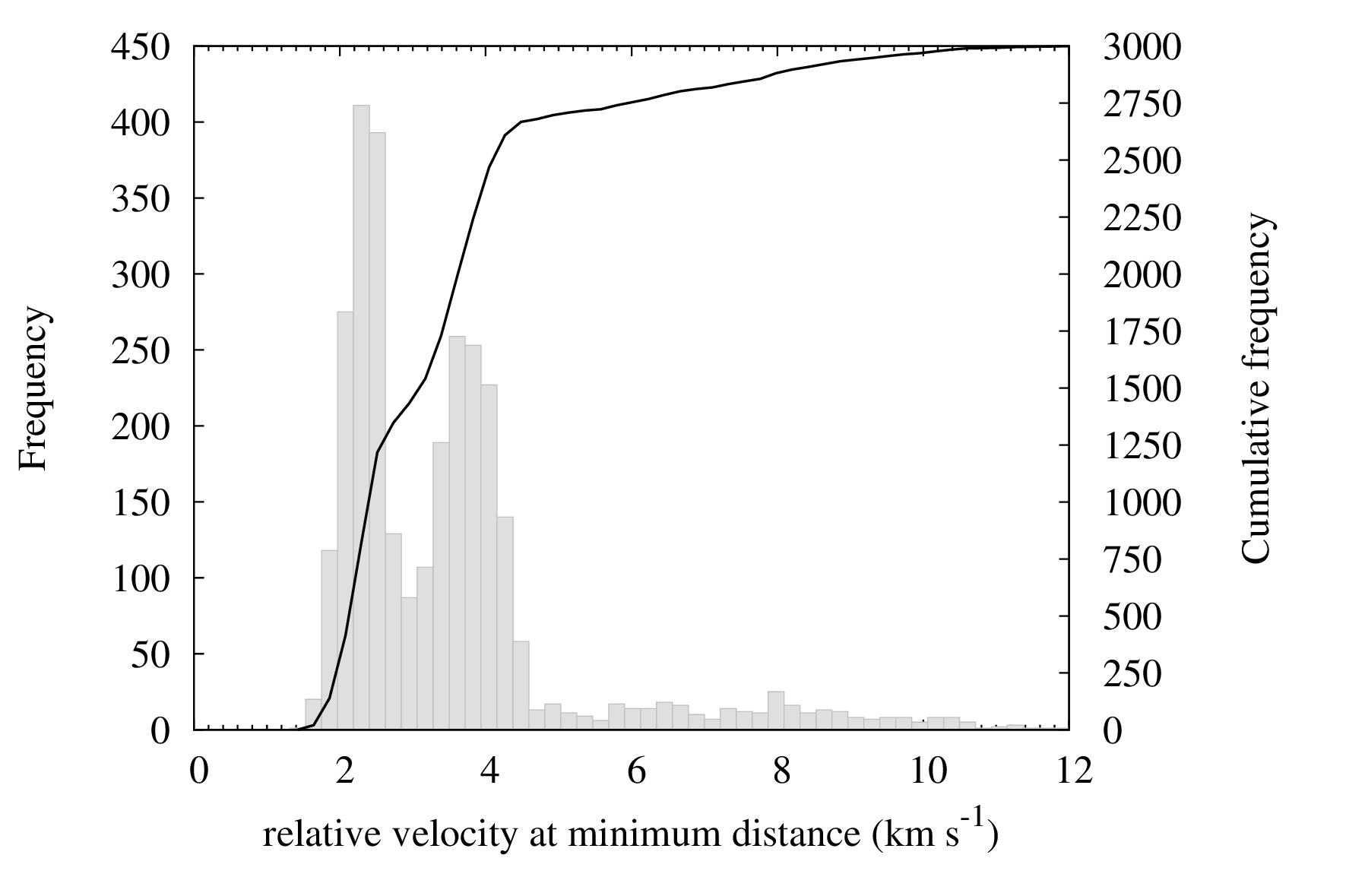}
         \caption{As Fig.~\ref{YB} but for control orbits statistically compatible with those of (54509)~YORP 2000~PH$_{5}$ and 
                  2007~WU$_{3}$. Here, the bin width is 0.2187~km~s$^{-1}$, with IQR = 1.5772~km~s$^{-1}$ (see the text for details). 
                 }
         \label{YW}
      \end{figure}

     The case of YORP and 2007~WU$_{3}$ (Gilmore et al. 2007; Bressi, Schwartz \& Holvorcem 2015) is markedly different from the previous 
     one and also from the bulk of pairs in Table \ref{yorps}, and deserves a more detailed consideration. A velocity distribution analysis 
     similar to that in Fig.~\ref{rvdist} but using control orbits for both objects generated by applying the covariance matrix methodology 
     described before and integrating backwards in time for 2\,000~yr (see Fig.~\ref{YW}) shows that the most probable (close to 80 per 
     cent) collision speed in this case is $<4$~km~s$^{-1}$ and over 20 per cent of encounters have relative velocities close to or below 
     2~km~s$^{-1}$. The velocity distribution is clearly bimodal with the most probable encounter velocities near the low-velocity extreme, 
     which is typical of encounters near aphelion where a multiplicity of flyby geometries are possible. The mean approach velocity falls in 
     the low-probability dip between the high-probability peak at about 2.3~km~s$^{-1}$ (encounters at aphelion) and the secondary peak at 
     about 3.8~km~s$^{-1}$ (encounters at perihelion). However, the closest approaches (at about 19\,000~km from each other) are 
     characterized by values of the relative velocity close to 2.5~km~s$^{-1}$. In addition, the probability of experiencing an encounter 
     closer than 30\,000~km is 0.0007. 

     Although the orbital solutions of both objects are not as precise as those of widely accepted young genetic pairs ---for example, 7343 
     Ockeghem (1992~GE$_{2}$) and 154634 (2003~XX$_{28}$) as studied by Duddy et al. (2012) or 87887 (2000~SS$_{286}$) and 415992 
     (2002~AT$_{49}$) as discussed by {\v Z}i{\v z}ka et al. (2016)--- and their dynamical environment is far more chaotic, this unusual 
     result hints at a possible physical connection between these two NEAs although perhaps the YORP mechanism was not involved in this 
     case. The velocity distributions for asteroid families studied by Bottke et al. (1994) ---see their figs 10a, 11a, 12a and 13a--- 
     clearly resemble what is observed in Fig.~\ref{YW}, in particular that of the Eos family (their fig 10a) which is non-Gaussian and 
     bimodal with the most probable collisions occurring near the apses; the lowest encounter velocities are observed when their perihelia 
     are aligned and the highest when they are anti-aligned. In the case of YORP and 2007~WU$_{3}$, low-velocity encounters take place at 
     about 1.22~au from the Sun and the high-velocity ones at 0.81~au. These two objects are no longer observable from the ground and they
     will remain at low solar elongation as observed from the Earth for many decades.
%
%
%
%

     The Eos family is the third most populous in the main asteroid belt and is thought to be the result of a catastrophic collision 
     (Hirayama 1918; Moth\'e-Diniz \& Carvano 2005), although its long-term evolution is also driven by the YORP effect (Vokrouhlick{\'y} et 
     al. 2006); many of its members are of the K spectral type and others resemble the S-type (see e.g. Moth\'e-Diniz, Roig \& Carvano 
     2005). Having a comparable short-term orbital evolution can be used to argue for a dynamical connection but not to claim a physical 
     connection. Genetic pairs must also have a similar chemical composition that can be studied via visible or near-infrared spectroscopy. 
     Gietzen \& Lacy (2007) carried out near-infrared spectroscopic observations of YORP and concluded that it belongs to either of the 
     silicaceous taxonomic classes S or V. Mueller (2007) has pointed out that a S-type classification would be in excellent agreement with 
     data from the Spitzer Space Telescope. Asteroid 2007~WU$_{3}$ may be of Sq or Q taxonomy according to preliminary results obtained by 
     the NEOShield-2 collaboration (Dotto et al. 
     2015).\footnote{http://www.neoshield.eu/wp-content/uploads/NEOShield-2\_D10.2\_i1\_\_Intermediate-observations-and-analysis-progress-report.pdf}
     If YORP and 2007~WU$_{3}$ are genetically related, their mutual short-term dynamical evolution suggests that they are the result of a
     catastrophic collision not the YORP effect. Such collisions are unusual, but not uncommon; asteroid (596)~Scheila 1906~UA experienced a 
     sub-critical impact in 2010 December by another main belt asteroid less than 100 m in diameter (Bodewits et al. 2011; Ishiguro et al. 
     2011; Jewitt et al. 2011; Moreno et al. 2011; Yang \& Hsieh 2011; Bodewits, Vincent \& Kelley 2014). In this case, the impact velocity 
     was probably close to 5~km~s$^{-1}$ (Ishiguro et al. 2011).

  \section{Encounters with planets and other NEOs}
     In addition to experiencing close encounters with other objects moving in YORP-like orbits, Fig. \ref{control} shows that these 
     interesting minor bodies can undergo close encounters with Venus, the Earth and (rarely) Mars. Close encounters with members of the 
     general NEO population are possible as well. The circumstances surrounding such encounters are explored statistically in this section.

     \subsection{Encountering other NEOs}
\begin{figure}
        \centering
         \includegraphics[width=\linewidth]{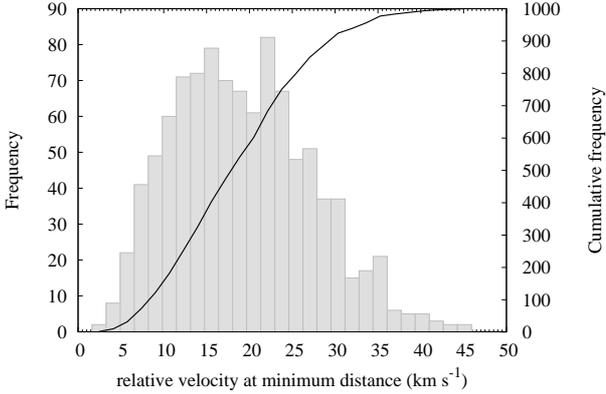}
         \caption{Velocity distribution between pairs of virtual NEAs, one of them following a YORP-like orbit and the second one a member 
                  of the general NEO population assuming uniformly distributed orbital elements (IQR = 8.2~km~s$^{-1}$, see the text for 
                  details).
                 }
         \label{vrNEA}
      \end{figure}

        We have already found that collisions between NEAs following YORP-like orbits are most likely happening at relative velocities close 
        to 13~km~s$^{-1}$, but most NEAs are not moving in YORP-like orbits. Here, we study the most general case of a close encounter 
        between a virtual NEA moving along a YORP-like trajectory and a member of the general NEO population. In order to explore properly 
        the available orbital parameter space, our population of virtual general NEAs has values of the orbital elements uniformly 
        distributed within the domain $q<1.3$~au (uniform in $q$, not in $a$), $e\in(0, 0.9)$, $i\in(0, 50)$\degr, $\Omega\in(0, 360)$\degr, 
        and $\omega\in(0, 360)$\degr. Fig. \ref{vrNEA} shows a velocity distribution that is very different from that in Fig. \ref{rvdist}; 
        instead of being nearly bimodal, it is clearly unimodal although the most likely value of the encounter velocity is similar to that
        in Fig. \ref{rvdist}. About 90 per cent of encounters/collisions have characteristic relative velocities in excess of 7~km~s$^{-1}$
        and about 55 per cent, above 15~km~s$^{-1}$. If one of the minor bodies discussed here experiences a collision with another small 
        NEA, the impact speed will be probably high enough to disrupt the object, partially or fully, creating a group of genetically (both 
        physically and dynamically) related bodies.  
%
%
%
%

\begin{figure}
        \centering
         \includegraphics[width=\linewidth]{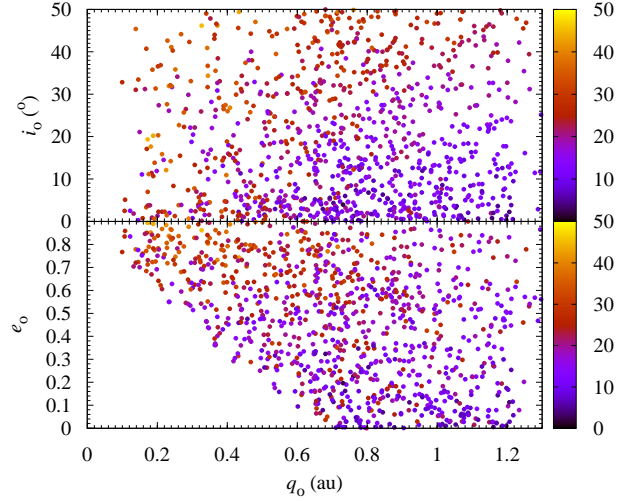}
         \caption{Velocity distribution as a function of the initial values of $q$ and $e$ (bottom panel) and $q$ and $i$ (top panel). The 
                  colours in the colour maps are proportional to the value of the velocity in Fig. \ref{vrNEA}.
                 }
         \label{mapvrNEA}
      \end{figure}

        Fig. \ref{mapvrNEA} shows the encounter velocity (colour coded) for the pairs in Fig. \ref{vrNEA} as a function of the initial 
        values of the orbital parameters ---$q$, $e$ and $i$--- of the general NEA. As NEAs with low values of $q$ and high values of $i$ 
        are rare, very high-speed collisions are unlikely. Relatively low-speed collisions mostly involve NEAs moving in low-eccentricity,
        low-inclination orbits. The lowest velocities ($<2.5$~km~s$^{-1}$) have been found for $a\in(0.9, 1.1)$~au, $e\in(0.2, 0.3)$, and 
        $i<4${\degr} which is precisely the volume of the orbital parameter space enclosing those NEAs following YORP-like orbits. As 
        expected, very low-speed collisions are only possible among members of the same dynamical class (i.e. when they have very similar 
        orbits).
%
%
%
%

\begin{figure}
        \centering
         \includegraphics[width=\linewidth]{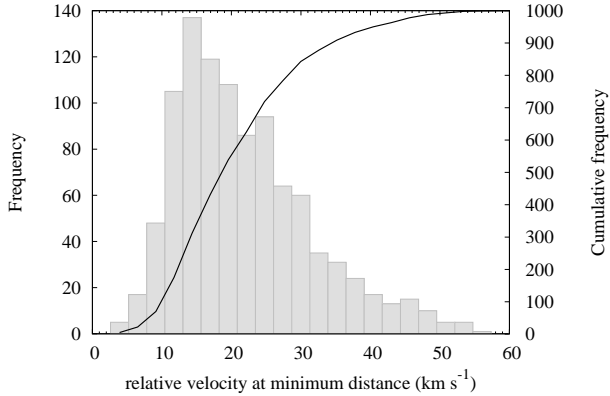}
         \caption{As Fig. \ref{vrNEA} but for encounters with a realistic population of synthetic NEOs from the NEOPOP model 
                  (IQR = 13.0~km~s$^{-1}$, see the text for details).
                 }
         \label{vrNEOPOP}
      \end{figure}

\begin{figure}
        \centering
         \includegraphics[width=\linewidth]{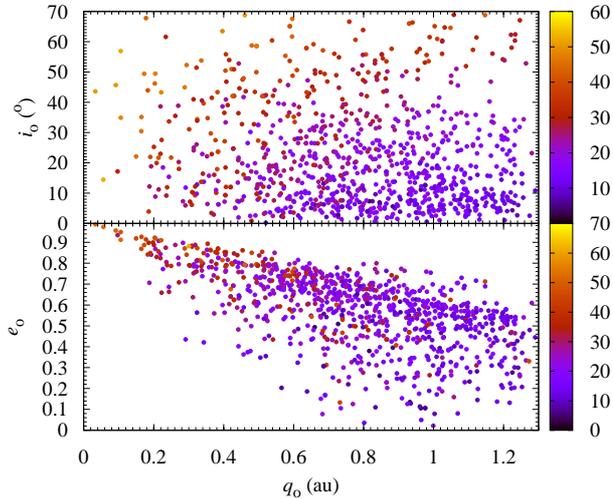}
         \caption{As Fig. \ref{mapvrNEA} but using data from the set of simulations in Fig. \ref{vrNEOPOP}.
                 }
         \label{mapvrNEOPOP}
      \end{figure}

        The previous analysis is based on the wrong assumption that NEOs have values of the orbital elements uniformly distributed; 
        Fig.~\ref{bias} clearly shows that this is not the case. On the other hand, sampling uniformly can oversample high eccentricities
        and inclinations, which lead to higher encounter velocities. In order to obtain unbiased estimates, we have performed additional 
        calculations using a population of synthetic NEOs from the NEOPOP model to obtain Figs \ref{vrNEOPOP} and \ref{mapvrNEOPOP}. The 
        most probable value of the relative velocity during encounters is $\sim$15~km~s$^{-1}$ and the most likely value of the semimajor 
        axis of the orbits followed by NEOs experiencing flybys with NEAs following YORP-like orbits is $\sim$2.5~au. This more realistic 
        analysis still shows that the most probable impact speed during a hypothetical collision between a general NEO and one object 
        following a YORP-like orbit could be high enough to cause significant damage to both NEOs.
%
%
%
%
%
%
%
%

     \subsection{Encountering Venus}
        Tables \ref{elements} and \ref{yorps}, and Figs \ref{control} and \ref{control2} show that minor bodies following YORP-like orbits 
        have perihelion distances that place them in the neighbourhood of the orbit of Venus. This fact clears the way to potential close 
        encounters with this planet as one of the nodes could be in the path of Venus (see Fig. \ref{control}, H-panels). In order to 
        investigate this possibility, we have performed short integrations (50 yr) of virtual NEAs with orbits as those described in 
        Section~3 and under the same conditions. We decided to use short integrations to minimize the effects on our results derived from 
        the inherently chaotic orbital evolution of these objects. Out of all the experiments performed, we have focused on 25\,000 cases 
        where the minimum distance between the virtual NEA and Venus during the simulation reached values under 0.1~au. We have found that 
        for this type of orbit, the probability of experiencing such close encounters with Venus is about 82.2 per cent (this value of the
        probability has been found dividing the number of experiments featuring encounters under 0.1~au by the total number of experiments 
        performed). Our calculations show that nearly 2 per cent of YORP-like orbits can lead to encounters under one Hill radius of Venus 
        (0.0067~au), about 0.03 per cent approach closer than 20 planetary radii and 0.01 per cent closer than 10 planetary radii. 

\begin{figure}
        \centering
        \includegraphics[width=\linewidth]{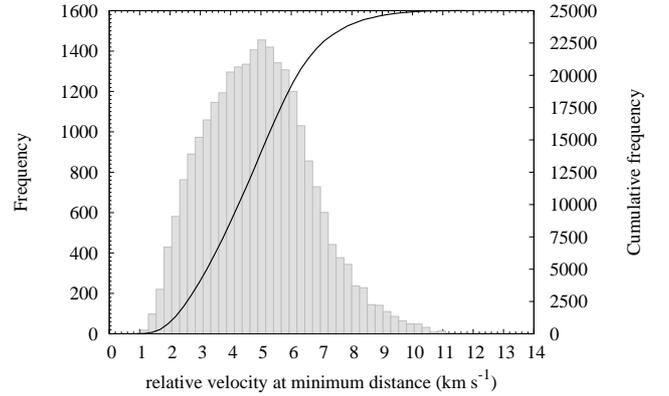}
        \caption{Distribution of relative velocities at closest approach for flybys with Venus of virtual NEAs following YORP-like orbits
                 (IQR = 2.407~km~s$^{-1}$).
                }
        \label{venuenc}
      \end{figure}

        Although relatively close encounters are frequent, the probability of placing the virtual NEA within the volume of space that could 
        be occupied by putative natural satellites of Venus is rather small for these orbits. Outside 10 planetary radii, the velocity of 
        the incoming body with respect to Venus during the closest encounters is in the range (5, 7)~km~s$^{-1}$; inside 10 planetary radii 
        the speed of the virtual NEA is significantly increased due to gravitational focusing by Venus. Fig. \ref{venuenc} shows that the 
        distribution of relative velocities at minimum distance is far from symmetric and the peak of the distribution does not correspond 
        to the typical values observed when the distance from Venus to the virtual NEA is the shortest possible. In principle, NEAs that 
        experience close encounters with Venus (or the Earth) can suffer resurfacing events when large rocks are dislodged from the surface 
        of the minor body. Our calculations indicate that this may be happening to NEAs following YORP-like orbits, but the chance of this 
        happening is low during encounters with Venus ---the probability of encounters under 10 Venus radii is just 0.01 per cent. 
%
%
%
%

     \subsection{Encountering the Earth: evaluating the impact risk}
        Here, we use the same set of simulations analysed in the previous section to perform a risk impact assessment evaluating the 
        probability that an object moving in a YORP-like orbit experiences a close encounter with our planet. We found that the probability 
        of suffering a close encounter under 0.1~au with the Earth is about 74.8 per cent (from 20\,000 experiments). Our calculations also 
        show that over 10 per cent of YORP-like orbits can approach the Earth closer than one Hill radius (0.0098~au), about 0.14 per cent 
        approach under 20 planetary radii and 0.05 per cent below 10 planetary radii. The overall upper limit for the impact probability 
        with our planet could be $< 0.00005$ which is higher than that with Venus. 

\begin{figure}
        \centering
         \includegraphics[width=\linewidth]{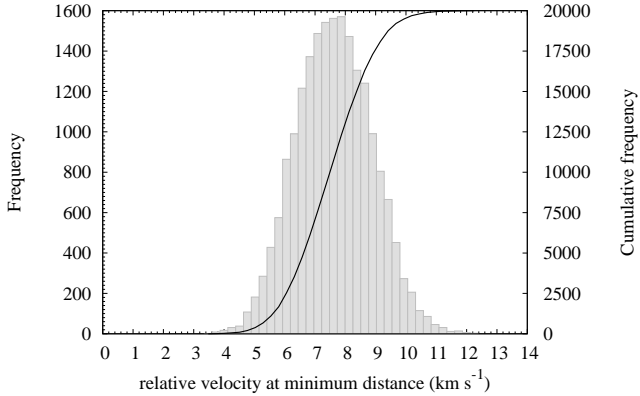}
         \caption{Similar to Fig. \ref{venuenc} but for encounters with the Earth. The velocity distribution is very approximately Gaussian 
                  with a most probable value of the relative velocity at closest approach of 7.631~km~s$^{-1}$ (IQR = 1.749~km~s$^{-1}$). 
                 }
         \label{earenc}
      \end{figure}

        Although the objects of interest here tend to approach Venus more frequently, when they approach our planet, they do it at a closer 
        distance. In principle, the Earth could be more efficient in altering the orbits of these small bodies, but Fig.~\ref{earenc} shows 
        that the most probable value of the relative velocity at closest approach is nearly 8~km~s$^{-1}$ with a Gaussian spread or standard 
        deviation of 1.230~km~s$^{-1}$. This is higher than for flybys with Venus. Encounters at lower relative speed could be more 
        effective at modifying the path of the perturbed minor body because it will spend more time under the influence of the perturber 
        during the flyby. Taking into account that the values of the masses of Venus and the Earth are very close, the overall influence of 
        both planets on the orbits of NEAs following YORP-like orbits could be fairly similar. In addition, close encounters can happen in
        sequence; i.e. an inbound NEO can experience a close flyby with our planet that may facilitate a subsequent close encounter with 
        Venus or an outbound NEO can approach Venus at close range making a close flyby with the Earth possible immediately afterwards.
%
%
%
%

        Figs~\ref{venuenc} and \ref{earenc} show that the most probable value of the relative velocity at closest approach is higher for 
        close encounters with our planet than for flybys with Venus and one may find this counterintuitive because at $\sim$0.7~au Keplerian
        velocities are higher than at $\sim$1~au. However, for flybys near perihelion multiple encounter geometries are possible. The 
        situation could be analogous to that of Amor asteroids (or Apollos with perihelion distances $\sim$1~au) encountering the Earth. 
        Independent calculations by JPL's \textsc{horizons} give a range of 4.94 to 11.64~km~s$^{-1}$ for the relative velocity during 
        encounters between 2017~FZ$_{2}$ and Venus during a time span of nearly 300 yr; in the case of encounters with the Earth it gives a 
        range of 4.63 to 22.52~km~s$^{-1}$.

        Several objects in Table~\ref{yorps} have very small values of their MOIDs with respect to the Earth, some as small as 30\,000 km or 
        under 5 Earth radii. NEAs moving in these orbits clearly pose a potential risk of impact with our planet. In addition to 
        2017~FZ$_{2}$ (and its initially possible impact in 2101--2104), two other small NEAs in Table \ref{yorps} have a non-negligible 
        probability of colliding with our planet in the near future as computed by JPL's Sentry System. Asteroid 2010~FN has an estimated 
        impact probability of 0.0000039 for a possible collision in 2079--2116.\footnote{https://cneos.jpl.nasa.gov/sentry/details.html\#?des=2010\%20FN} 
        Asteroid 2016~JA has an impact probability of 0.000001 for a possible impact in 
        2064--2103.\footnote{https://cneos.jpl.nasa.gov/sentry/details.html\#?des=2016\%20JA}  

     \subsection{Encountering Mars}
\begin{figure}
        \centering
         \includegraphics[width=\linewidth]{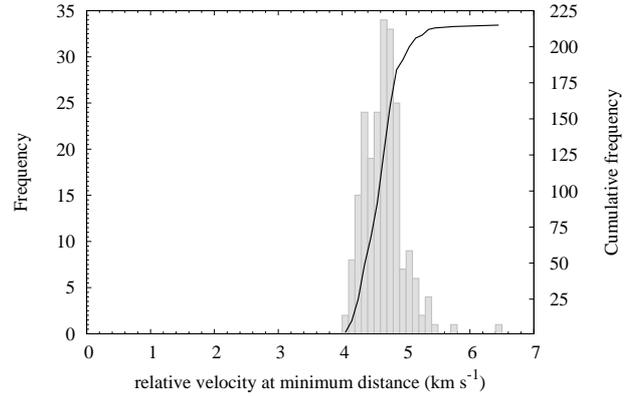}
         \caption{Similar to Fig. \ref{venuenc} but for encounters with Mars (IQR = 360.5~m~s$^{-1}$). 
                 }
         \label{marsenc}
      \end{figure}

        Calculations analogous to the ones described in the cases of Venus and the Earth show that although relatively distant encounters
        ($\sim$0.1~au) between objects moving in YORP-like orbits and Mars are possible, the actual probability is about 0.6 per cent and we 
        found no encounters under one Hill radius of Mars (0.0066~au). The velocity of these bodies relative to Mars at its smallest 
        separation is in the interval (4, 5)~km~s$^{-1}$ (see Fig. \ref{marsenc}). The role played by Mars on the dynamical evolution of 
        these minor bodies is clearly negligible.  
%
%
%
%

  \section{Discussion}
     Asteroid 2017~FZ$_{2}$ was, prior to its close encounter with the Earth on 2017 March 23, a quasi-satellite of our planet. It was the
     smallest detected so far ($H=26.7$~mag) and also moved in the least inclined orbit ($i=1\fdg7$). This minor body is no longer trapped
     in the 1:1 mean-motion resonance with our planet, but it has been trapped in the past and it could be again in the future. Its orbit is 
     highly chaotic and it was even suggested that this NEA may collide with our planet in the near future ($<100$~yr), but it is seen as of 
     less potential concern because of its small size. 

     Although apparently insignificant, this small body has led us to uncover a larger set of NEAs that appears to be at least some sort of 
     grouping of dynamical nature that perhaps includes several objects that may be genetically related. It is not possible to reach robust 
     conclusions regarding a possible connection between them because there is no available spectroscopic information for the vast majority 
     of these NEAs and their current orbital solutions are not yet sufficiently reliable. Most of these NEAs have few observations and/or 
     have not been re-observed for many years; a few of them, (54509)~YORP 2000~PH$_{5}$ included, may remain out of reach of ground-based 
     telescopes for many decades into the future.   

     Objects in YORP-like orbits\footnote{As pointed out above, $D_{\rm LS}$ and $D_{\rm R}<0.05$ with respect to YORP or (see 
     Table~\ref{yorps}, bottom section) $a\in(0.98, 1.02)$~au, $e\in(0.20, 0.30)$, $i\in(0, 4)$\degr, $\Omega\in(0, 360)$\degr, and 
     $\omega\in(0, 360)$\degr.} might be returning spacecraft, but their obvious excess makes this putative origin very unlikely. Although
     a number of items of space debris and working spacecraft (e.g. Rosetta, Gaia or Hayabusa 2) have been erroneously given minor body 
     designations, their orbits tend to be comparatively less inclined and far less eccentric, and their absolute magnitudes closer to 30. 
     In addition, any artificial interloper within the group of objects moving in YORP-like orbits would have been originally Venus-bound 
     and there are not many of those; besides, artificial objects without proper mission control input tend to drift away from Earth's 
     co-orbital region rather quickly (see e.g. section 9 of de la Fuente Marcos \& de la Fuente Marcos 2015b). They may be lunar ejecta, 
     but our calculations suggest that these objects cannot remain in their orbits for a sufficiently long period of time, say for the last 
     100\,000~yr. The only alternative locations for the origin of these bodies are either the asteroid belt or having been produced in situ 
     (i.e. within Earth's immediate neighbourhood) via YORP spin-up or any other mechanism able to generate fragments. Our analysis using 
     the NEOSSat-1.0 and NEOPOP orbital models indicates that the known delivery routes from the asteroid belt to the YORP-like orbital 
     realm might not be efficient enough to explain the high proportion of small NEAs currently moving in YORP-like orbits. The presence of 
     several relatively large minor bodies, YORP included, and a rather numerous dynamical cohort of much smaller objects hints at a 
     possible genetic relationship between several of the members of this group. Arguing for a genetic relationship requires both robust 
     dynamical and compositional similarities. However, it is also entirely possible that the existence of observational biases may account, 
     at least partially, for the observed excess.  

     Small NEAs could be the result of the YORP mechanism, but they may also be the by-product of collisions or the aftermath of tidal 
     stripping events during low-velocity close encounters with the planets, in particular with the Earth (see e.g. Bottke \& Melosh 
     1996a,b). Asteroids encountering the Earth (or Venus) with low relative velocity are good candidates for disruption because tidal 
     splitting is more likely at low relative velocity. Objects following YORP-like orbits encounter the Earth preferentially with relative 
     velocity in the range 6--8 km~s$^{-1}$ (the encounter statistics analysed in Section 5.3). Morbidelli et al. (2006) have shown that 
     (see their fig.~2), when encountering the Earth at these low relative velocities, any fragment resulting from tidal stripping can 
     become physically unbound in a relatively short time-scale, producing a pair of genetically related NEAs. 

\begin{figure}
        \centering
         \includegraphics[width=\linewidth]{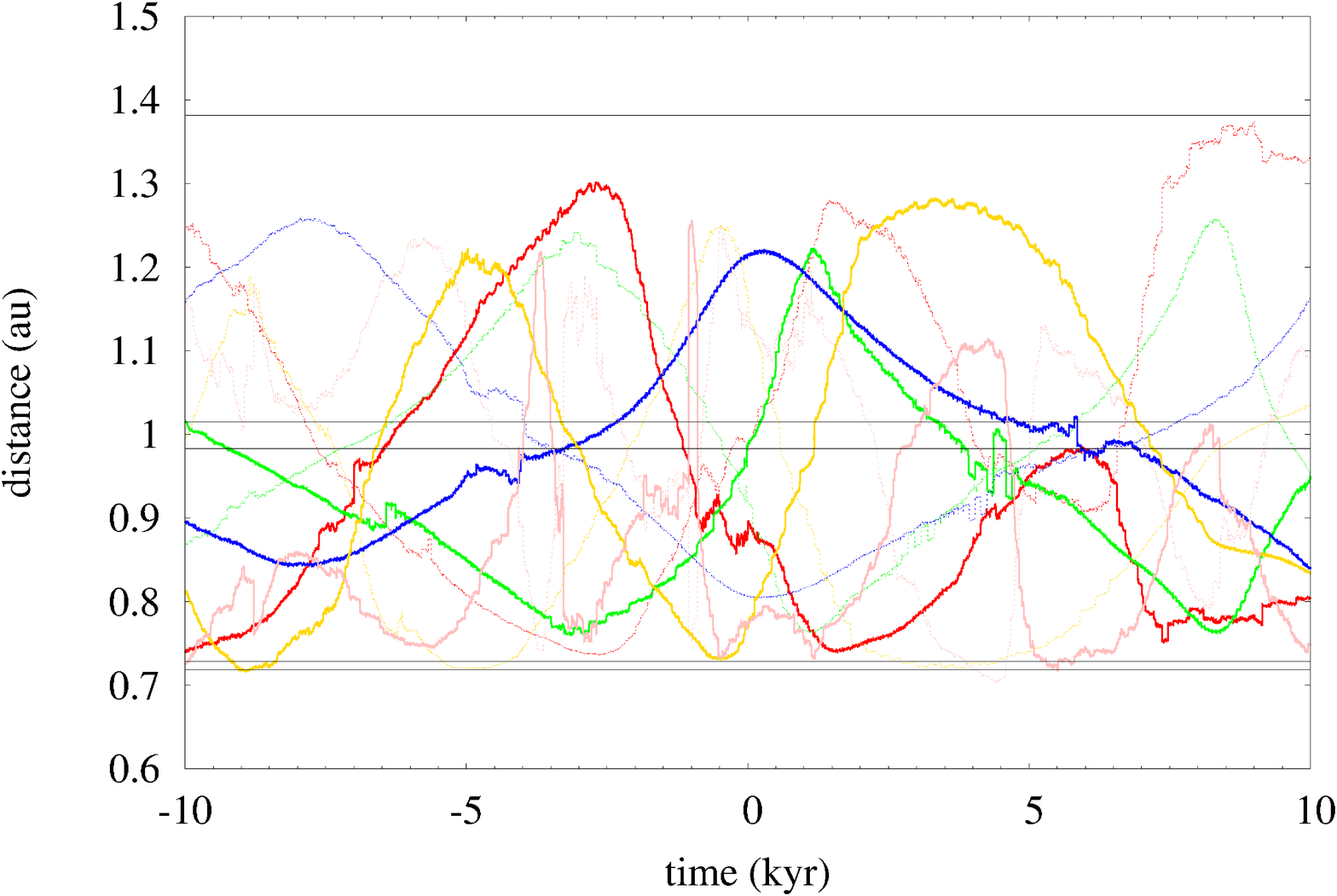}
         \caption{Evolution of the nodal distances for 2017~FZ$_{2}$ (red), (54509)~YORP 2000~PH$_{5}$ (green), 2007~WU$_{3}$ (blue), 
                  2009~HE$_{60}$ (gold) and 2012~VZ$_{19}$ (pink); epoch 2017 September 4. 
                 }
         \label{nodes}
      \end{figure}

     On the other hand, low-speed flybys, which are more sensitive to gravitational perturbations from our planet, tend to retain the 
     coherence of the nodal distances for shorter time-scales and are more prone to generate a dynamical grouping or meteoroid stream after 
     tidal disruption. Fig.~\ref{nodes} shows the behaviour of the nodal distances for a number of objects in Table~\ref{yorps} (also see 
     H-panels in Figs \ref{control} and \ref{control2}). As the orbit determinations of 2009~HE$_{60}$ and 2012~VZ$_{19}$ are rather poor, 
     Fig.~\ref{nodes} is only meant to be an example, not a numerically rigorous exploration of this important issue; in any case and 
     although certain objects appear to exhibit some degree of (brief) nodal coherence, the evolution is often too chaotic to be able to 
     arrive to any reliable conclusions. The pair-wise approach followed in Section~4 seems to be the only methodology capable of producing 
     robust evidence, either in favour or against a possible dynamical relationship.
%
%
%
%
 
     The picture that emerges from our extensive analysis is a rather complex one. YORP-induced rotational disruption may be behind the 
     origin of some of the small bodies discussed in this paper (as it could be the case of the pair YORP--2012~BD$_{14}$), but the fact 
     that they also experience very close encounters with both Venus and our planet cannot be neglected. On the other hand, once fragments 
     are produced either during planetary encounters or as a result of the YORP effect, these meteoroids can impact other objects moving 
     along similar paths at relatively high velocities triggering additional fragmentations (perhaps the case of the pair 
     YORP--2007~WU$_{3}$) or even catastrophic disruptions leading to some type of cascading effect that can eventually increase the size of 
     the population moving along these peculiar orbits and perhaps explain the observed excess. Subcatastrophic collisions and tidal 
     encounters can also lead to further rotational disruptions. The dynamical environment found here is rather different from the one 
     surrounding typical main-belt asteroid families; here, very close planetary encounters are possible in addition to the active role 
     played by mean-motion and secular resonances that are also present in the main asteroid belt.

     In this paper, we have not included the results of the Yarkovsky and YORP effects (see e.g. Bottke et al. 2006), but ignoring these 
     effects has no relevant impact on our analysis, which is based on relatively short integrations, or our conclusions. Not including 
     these non-gravitational forces in our integrations is justified by the fact that the largest predicted Yarkovsky drift rates are 
     $\sim$10$^{-7}$~au~yr$^{-1}$ (see e.g. Farnocchia et al. 2013) and also because the physical properties ---such as rotation rate, 
     albedo, bulk density, surface conductivity or emissivity--- of most of the NEAs under study here are not yet known and without them, no
     reliable calculations can be attempted. The width of the co-orbital region of the Earth is about 0.012~au; at the highest predicted 
     Yarkovsky drift rates, the characteristic time-scale to leak from the co-orbital region of our planet is equal to or longer than 
     120\,000~yr that is much larger than the time-scales pertinent to this study ---for an average value for the Yarkovsky drift of 
     10$^{-9}$~au~yr$^{-1}$, the time-scale to abandon the region of interest reaches 12 Myr.   
 
  \section{Conclusions}
     In this paper, we have studied the orbital evolution of the recently discovered NEA 2017~FZ$_{2}$ and other, perhaps related, minor 
     bodies. This research has been performed using $N$-body simulations and statistical analyses. Our conclusions can be summarized as 
     follows.
     \begin{enumerate}[(i)]
        \item Asteroid 2017~FZ$_{2}$ was until very recently an Earth's co-orbital, the sixth known quasi-satellite of our planet and the 
              smallest by far. Its most recent quasi-satellite episode may have started over 225~yr ago and certainly ended after a close 
              encounter with the Earth on 2017 March 23. 
        \item Extensive $N$-body simulations show that the orbit of 2017~FZ$_{2}$ is very unstable, with a Lyapunov time of the order of 
              100~yr. 
        \item Our orbital analysis shows that 2017~FZ$_{2}$ is a suitable candidate for being observed spectroscopically as it will remain 
              relatively well positioned with respect to our planet during its next visit in 2018. 
        \item Over a dozen other NEAs move in orbits similar to that of 2017~FZ$_{2}$, the largest named being (54509)~YORP 2000~PH$_{5}$. 
              Among these objects, we have identified two present-day co-orbitals of the Earth not previously documented in the literature: 
              2017~DR$_{109}$ follows a path of the horseshoe type and 2009~HE$_{60}$ is confirmed as a strong candidate to being a 
              quasi-satellite. 
        \item We find an apparent excess of small bodies moving in orbits akin to that of YORP that amounts to an over twentyfold increase 
              with respect to predictions from two different orbital models and we argue that this could be the result of mass shedding from 
              YORP.
        \item NEAs moving in YORP-like orbits may experience close encounters with both Venus and the Earth. Such encounters might lead to
              tidal disruption events, full or partial. We also find that mutual collisions are also possible within this group.
        \item $N$-body simulations in the form of extensive backwards integrations indicate that the pair YORP--2012~BD$_{14}$ might be the 
              recent ($\sim$10$^{3}$~yr) outcome of a YORP-induced rotational disruption.
        \item $N$-body simulations and the available spectroscopic evidence suggest that the pair YORP--2007~WU$_{3}$ may have a common
              genetic origin, perhaps a subcatastrophic collision.
     \end{enumerate}
     Spectroscopic studies during the next perigee of some of these objects should be able to confirm if YORP could be the source of any of
     the small NEAs studied here ---or if any of them is a relic of human space exploration.

  \section*{Acknowledgements}
     We thank the anonymous referee for a particularly constructive, detailed and very helpful first report and for additional comments, 
     S.~J. Aarseth for providing the code used in this research, A.~I. G\'omez de Castro, I. Lizasoain and L. Hern\'andez Y\'a\~nez of the 
     Universidad Complutense de Madrid (UCM) for providing access to computing facilities. This work was partially supported by the Spanish 
     `Ministerio de Econom\'{\i}a y Competitividad' (MINECO) under grant ESP2014-54243-R. Part of the calculations and the data analysis 
     were completed on the EOLO cluster of the UCM, and we thank S. Cano Als\'ua for his help during this stage. EOLO, the HPC of Climate 
     Change of the International Campus of Excellence of Moncloa, is funded by the MECD and MICINN. This is a contribution to the CEI 
     Moncloa. In preparation of this paper, we made use of the NASA Astrophysics Data System, the ASTRO-PH e-print server, and the MPC data 
     server.

  \bsp
  \label{lastpage}
\end{document}